\documentclass[10pt, conference]{IEEEtran}
\IEEEoverridecommandlockouts
\usepackage{cite}
\usepackage{amsmath,amssymb,amsfonts}
\usepackage{algorithmic}
\usepackage{graphicx}
\usepackage{textcomp}
\usepackage{xcolor}
\usepackage{comment}

\def\BibTeX{{\rm B\kern-.05em{\sc i\kern-.025em b}\kern-.08em
    T\kern-.1667em\lower.7ex\hbox{E}\kern-.125emX}}
\renewcommand\IEEEkeywordsname{Keywords}

\usepackage{amsthm}
\theoremstyle{plain}
\theoremstyle{definition}
\newtheorem{definition}{Definition}

\newtheorem{lemma}{Lemma}
\newtheorem{theorem}{Theorem}

\usepackage[titlenumbered,ruled,vlined,linesnumbered]{algorithm2e}

\usepackage{caption}
\def\tablename{Table}
\def\figurename{Figure}

\usepackage{wasysym}
\usepackage{subfigure}

\usepackage{tikz}
\usetikzlibrary{arrows,automata}
\usetikzlibrary{positioning}
\tikzset{
    state/.style={
           rectangle,
           rounded corners,
           draw=black, very thick,
           text centered,
           },
    fontscale/.style = {font=\relsize{#1}}
}

\usepackage{tikz-3dplot}
\usetikzlibrary{decorations.markings}
\usetikzlibrary{positioning}
\usetikzlibrary{calc}

\usepackage{hyperref}

\usepackage{makecell}
\usepackage{multirow}
\DeclareSymbolFont{largesymbolsA}{U}{txexa}{m}{n}
\DeclareMathSymbol{\varprod}{\mathop}{largesymbolsA}{16}

\newcommand{\crown}{$\alpha, \beta$-CROWN}

\begin{document}

\title{Expediting Neural Network Verification via Network Reduction}

\author{
\IEEEauthorblockN{Yuyi Zhong\textsuperscript{\textsection}}
\IEEEauthorblockA{\textit{School of Computing} \\
\textit{National University of Singapore}\\
Singapore \\
yuyizhong@u.nus.edu}
\and
\IEEEauthorblockN{Ruiwei Wang\textsuperscript{\textsection}\dag}
\IEEEauthorblockA{\textit{School of Computing} \\
\textit{National University of Singapore}\\
Singapore \\
wangruiwei@u.nus.edu}
\and
\IEEEauthorblockN{Siau-Cheng Khoo}
\IEEEauthorblockA{\textit{School of Computing} \\
\textit{National University of Singapore}\\
Singapore \\
khoosc@comp.nus.edu.sg}
}
\thispagestyle{plain}
\pagestyle{plain}
\maketitle
\begingroup\renewcommand\thefootnote{\textsection}
\footnotetext{Equal contribution}
\endgroup
\begingroup\renewcommand\thefootnote{\dag}
\footnotetext{Corresponding author}
\endgroup
\begin{abstract}
%
A wide range of verification methods have been proposed to verify the safety properties of deep neural networks ensuring that the networks function correctly in critical applications.
%
%
However, many well-known verification tools still struggle with complicated network architectures and large network sizes.
In this work, we propose a network reduction technique as a pre-processing method prior to verification.
The proposed method reduces neural networks via eliminating stable ReLU neurons, and transforming them into a sequential neural network consisting of ReLU and Affine layers which can be handled by the most verification tools.
%
%
We instantiate the reduction technique on the state-of-the-art complete and incomplete verification tools, including $\alpha$,$\beta$-crown, VeriNet and PRIMA.
%
Our experiments on a large set of benchmarks indicate that the proposed technique can significantly reduce neural networks and speed up existing verification tools.  
%
%
Furthermore, the experiment results also show that network reduction can improve the availability of existing verification tools on many networks by reducing them into sequential neural networks.

\end{abstract}

\vspace{-1em}
\section{Introduction}
Deep neural networks have been widely applied in real-world applications.
At the same time, it is indispensable to guarantee the safety properties of neural networks in those critical scenarios.
As neural networks are trained to be larger and deeper, researchers have deployed various techniques to speed up the verification process.
For example, to over-approximate the whole network behavior \cite{DBLP:conf/sp/GehrMDTCV18, singh2018fast, DBLP:journals/corr/abs-2201-01978, DBLP:conf/tacas/YangLLHWSXZ21}; to deploy GPU implementation \cite{DBLP:conf/iclr/FerrariMJV22, DBLP:journals/corr/abs-2103-06624, DBLP:conf/iclr/XuZ0WJLH21}; or to merge neurons in the same layer in an over-approximate manner so as to reduce the number of neurons \cite{DBLP:conf/sas/SotoudehT20, DBLP:conf/nips/PrabhakarA19}.

This work aims to further accelerate the verification process by ``pre-processing'' the tested neural network with ReLU activation function and constructing a \emph{reduced} network with \emph{fewer} number of neurons and connections. 
We propose the {\em network reduction} technique, which returns a reduced network (named as \emph{REDNet}) that captures the \emph{exact} behavior of the original network rather than over-approximating the original network.
Therefore verification over the reduced network equals the original verification problem yet requires less execution cost.

The REDNet could be instantiated on different verification techniques and is beneficial for complex verification processes such as branch-and-bound based (bab) or abstract refinement based methods.
For example, branch-and-bound based methods \cite{DBLP:journals/jmlr/BunelLTTKK20, DBLP:conf/iclr/LuK20, DBLP:journals/corr/abs-2103-06624, DBLP:conf/iclr/XuZ0WJLH21} generate a set of sub-problems to verify the original problem.
Once deployed with REDNet before the branch-and-bound phase, all the sub-problems are built on the reduced network, thus achieving overall speed gain.
For abstract refinement based methods, like \cite{DBLP:journals/pacmpl/MullerMSPV22, DBLP:conf/tacas/YangLLHWSXZ21, DBLP:conf/vmcai/ZhongTK23}, they collect and encode the network constraints and refine individual neuron bounds via LP (linear program) or MILP (mixed-integer linear program) solving.
This refinement process could be applied to a large set of neurons, and the number of constraints in the network encoding can be significantly reduced with the deployment of REDNet. 

We have implemented our proposed network reduction technique in a prototypical system named REDNet (the \underline{red}uced \underline{net}work), which is available at \url{https://github.com/REDNet-verifier/IDNN}.
The experiments show that the ReLU neurons in the reduced network could be up to 95 times smaller than those in the original network in the best case and 10.6 times smaller on average.
We instantiated REDNet on the state-of-the-art complete verifier \crown{} \cite{AlphaBetaCrownSystem}, VeriNet \cite{VerinetSystem} and incomplete verifier PRIMA \cite{DBLP:journals/pacmpl/MullerMSPV22} and assessed the effectiveness of REDNet over a wide range of benchmarks.
The results show that, with the deployment of REDNet, \crown{} could verify more properties given the same timeout and gain average $1.5\times$ speedup.
Also, VeriNet with REDNet verifies 25.9\% more properties than the original and can be $1.6\times$ faster.
REDNet could also assist PRIMA to gain $1.99 \times$ speedup and verifies 60.6\% more images on average.
Lastly, REDNet is constructed in a simple network architecture and making it amenable for existing tools to handle network architectures that they could not support previously.

We summarize the contributions of our work as follows:
\begin{itemize}
    \item We define stable ReLU neurons and deploy the state-of-the-art bounding method to detect such stable neurons. 
    \item We propose a \emph{network reduction} process to \emph{remove} stable neurons and generate REDNet that contains a smaller number of neurons and connections, thereby boosting the efficiency of existing verification methods.
    %
    \item We prove that the generated REDNet preserves the input-output equivalence of the original network. 
    \item We instantiate the REDNet on several state-of-the-art verification methods. 
    %
    The experiment results indicate that the same verification processes execute faster on the REDNet than on the original network; it can also respond accurately to tougher queries the verification of which were timeout when running on the original network. 
    \item REDNet is constructed with a simple network architecture, it can assist various tools in handling more networks that they have failed to be supported previously.
    \item Lastly, we export REDNet as a fully-connected network in ONNX, which is an open format that is widely accepted by the most verification tools.
\end{itemize}

%
%
%

\section{Preliminaries}
\vspace{-0.5em}
A {\em feedforward neural network} is a function $\mathcal{F}$ defined as a directed acyclic diagram $(\mathcal{V}, \mathcal{E})$ where every node $L_i$ in $\mathcal{V}$ represents a layer of $|L_i|$ neurons 
and each arc $(L_i, L_j)$ in $\mathcal{E}$ denotes that the outputs of the neurons in $L_i$ are inputs of the neurons in $L_j$.
For each layer $L_i\in \mathcal{V}$,  $in(L_i)=\{L_j|(L_j, L_i)\in \mathcal{E}\}$ is the {\em preceding layers} of $L_i$ and $out(L_i)=\{L_j|(L_i, L_j)\in \mathcal{E}\}$ denotes the {\em succeeding layers} of $L_i$. 
If the set $in(L_i)$ of preceding layers is non-empty, then the layer $L_i$ represents a computation operator, e.g. the ReLU and GEMM operators; otherwise $L_i$ is an input layer of the neural network.
In addition, $L_i$ is an {\em output layer} of the neural network if $out(L_i)$ is empty.

In this paper, we consider the ReLU-based neural network with one input layer and one output layer.
Note that multiple input layers (and output layers) can be concatenated as one input layer (and one output layer).
Then a neural network is considered as a {\em sequential neural network} if $|in(L_i)|=1$ and $|out(L_i)|=1$ for all layers $L_i$ in the neural network except the input and output layers.

We use a vector $\vec{a}$ to denote the input of a neural network $\mathcal{F}$, and the {\em input space} $I$ of $\mathcal{F}$ includes all possible inputs of $\mathcal{F}$.
For any input $\vec{a}\in I$, $L_i(\vec{a})$ is a vector denoting the outputs of neurons in a layer $L_i$ given this input $\vec{a}$.
The output $\mathcal{F}(\vec{a})$ of the neural network is the output of its output layer.
%

The {\em neural network verification problem} is to verify that for all possible inputs $\vec{a}$ from a given input space $I$, the outputs $\mathcal{F}(\vec{a})$ of a neural network $\mathcal{F}$ must satisfy a specific condition.
For example, the robustness verification problem is to verify that for all inputs within a specified input space, the neural network's output must satisfy that the value of the neuron corresponding to the ground truth label is greater than the values of the other neurons in the output layer.

\section{Overview}
\label{sec:REDoverview}
In this section, we present a simple network $\mathcal{F}$ and illustrate how to construct the reduced network via the deletion of stable neurons, where stable neurons refer to those ReLU neurons whose inputs are completely non-negative or non-positive.

The example is a fully-connected network with ReLU activation function $y = \text{max}(0,x)$ as shown in \autoref{fig:redEgNet}, where the connections are recorded in the linear constraints near each affine neuron. 
We set the input space $I$ to be $[-1,1] \times [-1, 1]$, and apply one of the state-of-the-art bound propagators CROWN \cite{DBLP:conf/nips/ZhangWCHD18} to propagate the input intervals to other neurons of the network.
The deployment of CROWN returns us the concrete bounds for intermediate neurons, which are displayed next to the corresponding neurons in \autoref{fig:redEgNet}. 

From this initial computation, we observe that four ReLU neurons are \emph{stable}: $x_8$ is stably deactivated as its input $x_3 \leq 0$ yields $x_8 = 0$; $x_9, x_{10}, x_{11}$ are stably activated as their inputs are always greater or equal to zero, yielding $x_9 = x_4, x_{10} = x_5, x_{11} = x_6$.
Given the observation, we could \emph{remove} those stable ReLU neurons together with their input neurons: we could directly eliminate neurons $x_3, x_8$; and delete $x_4, x_5, x_6, x_9, x_{10}, x_{11}$ while \emph{connecting} $y_1, y_2$ directly to the preceding neurons of $x_4, x_5, x_6$, which are $x_1, x_2$.

\begin{figure}[!t]
\scriptsize
\centering
\begin{tikzpicture}[
    red_node/.style={circle, draw=red, fill=red!5, thin,
      minimum size = 6mm, inner sep=1pt},
    blue_node/.style={circle, draw = blue, fill=cyan!5, thin,
      minimum size = 6mm, inner sep=1pt},
    black_node/.style={circle, draw = black, fill=black!5, thin,
      minimum size = 6mm, inner sep=1pt},
    red_rectangle/.style={rectangle, draw = red, thin, dashed}
    ]
  \node[red, red_node] (x1){$x_1$};
  \node[red, left = 10mm of x1] (interval1) {};
  \node[black, above left = -2mm and 0.5mm of x1] () {$[-1,1]$};
  \node[red, red_node, below = 11mm of x1](x2){$x_2$};
  \node[black, above left = -2mm and 0.5mm of x2] () {$[-1,1]$};
  \node[red, left = 10mm of x2] (interval2) {};
  \node[black, black_node, above right = 5mm and 15mm of x1] (x4){$x_4$};
 \node[black, black_node, above = 11mm of x4] (x3){$x_3$};
 \node[black, black_node, above right = 5mm and 15mm of x2] (x5){$x_5$};
 \node[black, black_node, below = 10mm of x5] (x6){$x_6$};
  \node[black, black_node, below = 10mm of x6] (x7){$x_7$};
\node[black, black_node, right = 16mm of x4] (r4){$x_9$};
 \node[black, black_node, right = 16mm of x3] (r3){$x_8$};
 \node[black, black_node, right = 16mm of x5] (r5){$x_{10}$};
 \node[black, black_node, right = 16mm of x6] (r6){$x_{11}$};
  \node[black, black_node, right = 16mm of x7] (r7){$x_{12}$};
 \node[blue, blue_node, right = 54mm of x1] (y1){$y_1$};
 \node[blue, blue_node, right = 54mm of x2] (y2){$y_2$};
\node (C0) at ($(x3)!0.5!(r3)$) {};
  \node (C1) at ($(x4)!0.5!(r4)$) {};
  \node (C2) at ($(x5)!0.5!(r5)$) {};
   \node (C3) at ($(x6)!0.5!(r6)$) {};
   \node (C4) at ($(x7)!0.5!(r7)$) {};
  \node[black, above = -1.5 mm of C0] () {$\mathrm{max}(0, x_3)$};
  \node[black, above = -1.5 mm of C1] () {$\mathrm{max}(0, x_4)$};
  \node[black, above = -1.5 mm of C2] () {$\mathrm{max}(0, x_5)$};
  \node[black, above = -1.5 mm of C3] () {$\mathrm{max}(0, x_6)$};
  \node[black, above = -1.5 mm of C4] () {$\mathrm{max}(0, x_7)$};
 \node[black, above = 0 mm of x3] (x3cub) {$x_3 \in [-4, 0]$};
 \node[black, above = -1 mm of x3cub] (x3def) {$-x_1-x_2-2$};
 \node[black, above right = 1 mm and -5mm of x4] (x4cub) {$x_4 \in [1,5]$};
 \node[black, above = -1 mm of x4cub] (x4def) {$x_1+x_2+3$};
 \node[black, above right = 1 mm and -7mm of x5] (x5cub) {$x_5 \in [0,4]$};
 \node[black, above = -1 mm of x5cub] (x5def) {$x_1-x_2+2$};
 \node[black, above right = 2 mm and -7mm of x6] (x6cub) {$x_6 \in[0,4]$};
  \node[black, above = -1 mm of x6cub] (x6def) {$x_1+x_2+2$};
 \node[black, above right = 2 mm and -5.5mm of x7] (x7cub) {$x_7 \in[-2,2]$};
  \node[black, above left = -1 mm and -13mm of x7cub] (x7def) {$-x_1+x_2$};
 \node[black, above = 0 mm of r3] (r3cub) {$x_8 \in [0,0]$};
 \node[black, above right = 1 mm and -11mm of r4] (r4cub) {$x_9 \in [1, 5]$};
 \node[black, above = 0 mm of r5] (r5cub) {$x_{10} \in [0,4]$};
\node[black, above right = 3 mm and -12mm of r6] (r6cub) {$x_{11} \in [0,4]$};
\node[black, above right = 5 mm and -12.5mm of r7] (r7cub) {$x_{12} \in [0,2]$};
\node[black, above right = 3 mm and -7mm of y1] (y1def) {$x_{11} - x_{12}$};
\node[black, above = 1 mm of y1def] (y1deff) {$x_8 - x_9 + x_{10}+$};
\node[black, below = 3 mm of y2] (y2def) {$x_8 + x_9 +$};
\node[black, below = 1 mm of y2def] (y2deff) {$x_{10} + x_{11} + x_{12}$};
  \draw [->,black,thin](interval1) -- (x1);
  \draw [->,black,thin](interval2) -- (x2);
  \draw [->,black,thin](x3) -- (r3);
  \draw [->,black,thin](x4) -- (r4);
  \draw [->,black,thin](x5) -- (r5);
  \draw [->,black,thin](x6) -- (r6);
  \draw [->,black,thin](x7) -- (r7);
  \draw [->,black,thin](x1) -- (x3);
  \draw [->,black,thin](x1) -- (x4);
  \draw [->,black,thin](x1) -- (x5);
  \draw [->,black,thin](x1) -- (x6);
  \draw [->,black,thin](x1) -- (x7);
  \draw [->,black,thin](x2) -- (x3);
  \draw [->,black,thin](x2) -- (x4);
  \draw [->,black,thin](x2) -- (x5);
  \draw [->,black,thin](x2) -- (x6);
  \draw [->,black,thin](x2) -- (x7);
  \draw [->,black,thin](r3) -- (y1);
  \draw [->,black,thin](r4) -- (y1);
  \draw [->,black,thin](r5) -- (y1);
  \draw [->,black,thin](r6) -- (y1);
  \draw [->,black,thin](r7) -- (y1);
  \draw [->,black,thin](r3) -- (y2);
  \draw [->,black,thin](r4) -- (y2);
  \draw [->,black,thin](r5) -- (y2);
  \draw [->,black,thin](r6) -- (y2);
  \draw [->,black,thin](r7) -- (y2);
\end{tikzpicture}
\vspace{-0.5em}
\caption{The example network with initial concrete bounds}
\label{fig:redEgNet}
\end{figure}
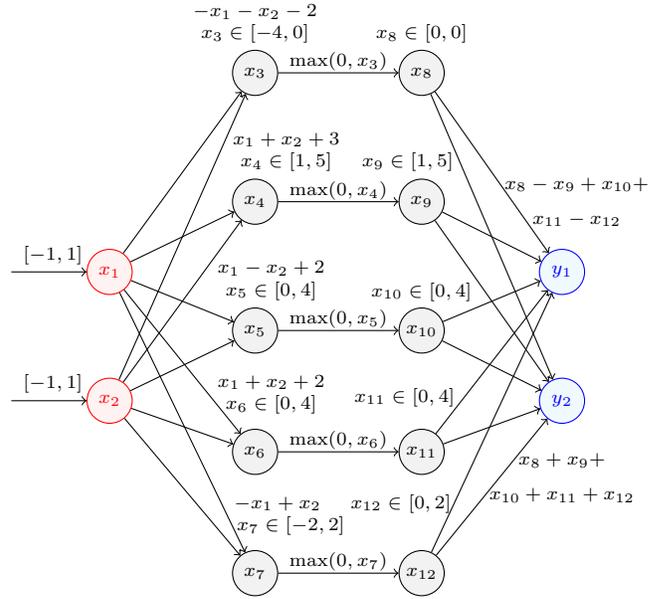
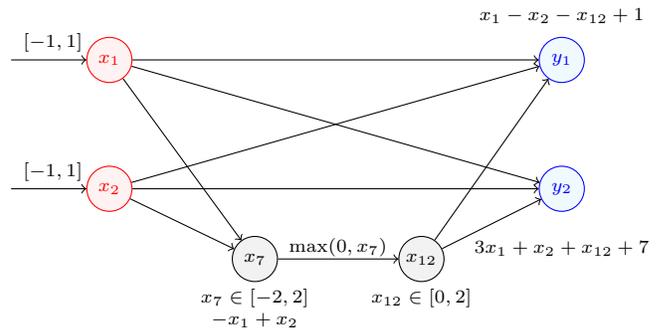
\begin{figure}[!b]
\scriptsize
\centering
\begin{tikzpicture}[
    red_node/.style={circle, draw=red, fill=red!5, thin,
      minimum size = 6mm, inner sep=1pt},
    blue_node/.style={circle, draw = blue, fill=cyan!5, thin,
      minimum size = 6mm, inner sep=1pt},
    black_node/.style={circle, draw = black, fill=black!5, thin,
      minimum size = 6mm, inner sep=1pt},
    red_rectangle/.style={rectangle, draw = red, thin, dashed}
    ]
  \node[red, red_node] (x1){$x_1$};
  \node[red, left = 10mm of x1] (interval1) {};
  \node[black, above left = -2mm and 0.5mm of x1] () {$[-1,1]$};
  \node[red, red_node, below = 11mm of x1](x2){$x_2$};
  \node[black, above left = -2mm and 0.5mm of x2] () {$[-1,1]$};
  \node[red, left = 10mm of x2] (interval2) {};
  \node[black, black_node, below right = 5mm and 15mm of x2] (x7){$x_7$};
  \node[black, black_node, right = 16mm of x7] (r7){$x_{12}$};
 \node[blue, blue_node, right = 54mm of x1] (y1){$y_1$};
 \node[blue, blue_node, right = 54mm of x2] (y2){$y_2$};
   \node (C4) at ($(x7)!0.5!(r7)$) {};
  \node[black, above = -1.5 mm of C4] () {$\mathrm{max}(0, x_7)$};
 \node[black, below  = 0 mm of x7] (x7cub) {$x_7 \in[-2,2]$};
  \node[black, below  = -1 mm of x7cub] (x7def) {$-x_1+x_2$};
\node[black, below  = 0mm of r7] (r7cub) {$x_{12} \in [0,2]$};
\node[black, above = 1 mm of y1] (y1def) {$x_1 - x_2 - x_{12} + 1$};
\node[black, below = 3 mm of y2] (y2def) {$3x_1 +x_2 + x_{12} + 7$};
  \draw [->,black,thin](interval1) -- (x1);
  \draw [->,black,thin](interval2) -- (x2);
  \draw [->,black,thin](x7) -- (r7);
  \draw [->,black,thin](x1) -- (y1);
  \draw [->,black,thin](x1) -- (y2);
  \draw [->,black,thin](x1) -- (x7);
  \draw [->,black,thin](x2) -- (y1);
  \draw [->,black,thin](x2) -- (y2);
  \draw [->,black,thin](x2) -- (x7);
  \draw [->,black,thin](r7) -- (y1);
  \draw [->,black,thin](r7) -- (y2);
\end{tikzpicture}
\vspace{-0.5em}
\caption{The network connections after neuron removal}
\label{fig:removalNet}
\end{figure}
After removal, the connections are updated as in \autoref{fig:removalNet}.
The new affine constraint of $y_1$ is computed as follows:
\vspace{-0.6em}
 \begin{equation}
y_1
\begin{array}[t]{ll}
= & x_8 - x_9 + x_{10} + x_{11} - x_{12}\\
= & 0 - x_4 + x_5 + x_6 - x_{12} \\
= & x_1 - x_2 + 1 - x_{12}
\end{array}
\end{equation}
\vspace{-0.2em}
Similarly, the computation of $y_2$ is updated as follows:
\vspace{-0.5em}
 \begin{equation}
y_2
\begin{array}[t]{ll}
= & x_8 + x_9 + x_{10} + x_{11} + x_{12}\\
= & 0 + x_4 + x_5 + x_6 + x_{12} \\
= & 3x_1 + x_2 + 7 + x_{12}
\end{array}
\end{equation}
The above computation only involves equality replacement; therefore, the two networks in \autoref{fig:removalNet} and \autoref{fig:redEgNet} functions \emph{the same} given the specified input space $I$.
However, the network architecture has been modified, and the output neurons are now defined over its preceding layer together with the input layer.
To preserve the network architecture without introducing new connections between layers, we \emph{merge} the stably activated neurons into \emph{a smaller set} of neurons instead of directly deleting them.

\begin{figure}[!t]
\scriptsize
\centering
\begin{tikzpicture}[
    red_node/.style={circle, draw=red, fill=red!5, thin,
      minimum size = 6mm, inner sep=1pt},
    blue_node/.style={circle, draw = blue, fill=cyan!5, thin,
      minimum size = 6mm, inner sep=1pt},
    black_node/.style={circle, draw = black, fill=black!5, thin,
      minimum size = 6mm, inner sep=1pt},
    red_rectangle/.style={rectangle, draw = red, thin, dashed}
    ]
  \node[red, red_node] (x1){$x_1$};
  \node[red, left = 10mm of x1] (interval1) {};
  \node[black, above left = -2mm and 0.5mm of x1] () {$[-1,1]$};
  \node[red, red_node, below = 11mm of x1](x2){$x_2$};
  \node[black, above left = -2mm and 0.5mm of x2] () {$[-1,1]$};
  \node[red, left = 10mm of x2] (interval2) {};
  \node[black, black_node, below right = 5mm and 15mm of x2] (x7){$x_7$};
  \node[black, black_node, above = 10mm of x7] (m2){$m_2$};
     \node[black, above right = 1 mm and -7mm of m2] (m2bounds) {$m_2 \in [3,11]$};
 \node[black, above = -1 mm of m2bounds] (m2def) {$3x_1+x_2+7$};
  \node[black, black_node, right = 16mm of m2] (m4){$m_4$};
   \node[black, above  = 1 mm of m4] (m4bounds) {$m_4 \in [3,11]$};
  \node[black, black_node, above = 11mm of m2] (m1){$m_1$};
   \node[black, above right = 1 mm and -7mm of m1] (m1bounds) {$m_1 \in [0,4]$};
 \node[black, above = -1 mm of m1bounds] (m1def) {$x_1-x_2+2$};
  \node[black, black_node, right = 16mm of m1] (m3){$m_3$};
 \node[black, above  = 1 mm of m3] (m3bounds) {$m_3 \in [0,4]$};
  \node[black, black_node, right = 16mm of x7] (r7){$x_{12}$};
 \node[blue, blue_node, right = 54mm of x1] (y1){$y_1$};
 \node[blue, blue_node, right = 54mm of x2] (y2){$y_2$};
  \node (C4) at ($(x7)!0.5!(r7)$) {};
  \node[black, above = -1.5 mm of C4] () {$\mathrm{max}(0, x_7)$};
  \node (C3) at ($(m1)!0.5!(m3)$) {};
  \node[black, above = -1.5 mm of C3] () {$\mathrm{max}(0, m_1)$};
  \node (C2) at ($(m2)!0.5!(m4)$) {};
  \node[black, above = -1.5 mm of C2] () {$\mathrm{max}(0, m_2)$};
 \node[black, below  = 0 mm of x7] (x7cub) {$x_7 \in[-2,2]$};
  \node[black, below  = -1 mm of x7cub] (x7def) {$-x_1+x_2$};
\node[black, below  = 0mm of r7] (r7cub) {$x_{12} \in [0,2]$};
\node[black, above = 2 mm of y1] (y1def) {$m_3 - x_{12} - 1$};
\node[black, below = 3 mm of y2] (y2def) {$m_4 + x_{12}$};
  \draw [->,black,thin](interval1) -- (x1);
  \draw [->,black,thin](interval2) -- (x2);
  \draw [->,black,thin](x7) -- (r7);
  \draw [->,black,thin](m1) -- (m3);
  \draw [->,black,thin](m2) -- (m4);
  \draw [->,black,thin](x1) -- (m1);
  \draw [->,black,thin](x1) -- (m2);
  \draw [->,black,thin](x1) -- (x7);
  \draw [->,black,thin](x2) -- (m1);
  \draw [->,black,thin](x2) -- (m2);
  \draw [->,black,thin](x2) -- (x7);
  \draw [->, black,thin](m3) -- (y1);
  \draw [->, dashed, black,thin](m4) -- (y1);
  \draw [->,black,thin](r7) -- (y1);
  \draw [->,black,dashed,thin](m3) -- (y2);
  \draw [->,black,thin](m4) -- (y2);
  \draw [->,black,thin](r7) -- (y2);
\end{tikzpicture}
\vspace{-0.5em}
\caption{The final network after reduction (REDNet), where the dashed connection means the coefficient equals to 0}
\label{fig:finalredNet}
\end{figure}
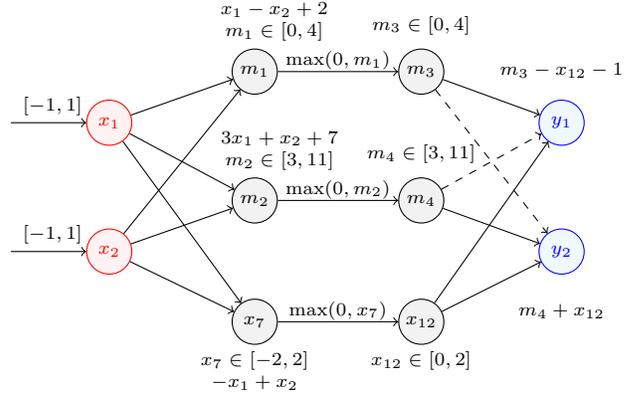

The final reduced network is shown below in \autoref{fig:finalredNet}, where we transform the connection between $y_1, y_2$ and $x_1, x_2$ in \autoref{fig:removalNet} into two merged neurons $m_1 = x_1-x_2 + 1; m_2 = 3x_1+x2 + 7$.
Since $m_2$ is stably activated given the input space, we have $m_4 = m_2$, thus $y_2 = m_4 + x_{12}$ which equals to the definition of $y_2$ in \autoref{fig:removalNet}.
To further enforce $m_1$ to remain stably activated, we increase the bias of $m_1$ to be 2, which leads to $m_3=m_1$, thus $y_2=m_3 -x_{12} -1$.
Therefore, the final reduced network in \autoref{fig:finalredNet} remains to be a fully-connected network, but the stably deactivated neuron $x_3$ has been removed, and the original set of stably activated neurons $x_4, x_5, x_6$ are merged into a smaller set of stably activated neurons $m_1, m_2$.
Note that the connection between $y_1, y_2$ and $m_1, m_2$ are actually identity matrix:
\begin{equation}
    \begin{bmatrix}
       y_1 \\
       y_2 \\
     \end{bmatrix} 
     =
     \begin{bmatrix}
       1 & 0\\
       0 & 1 \\
     \end{bmatrix} 
     \cdot
     \begin{bmatrix}
       m_1 \\
       m_2 \\
     \end{bmatrix} 
     +
     \begin{bmatrix}
       -1 \\
       1 \\
     \end{bmatrix} \cdot x_{12}
     +
    \begin{bmatrix}
       -1 \\
       0 \\
     \end{bmatrix}
\end{equation}
Therefore, the number of merged neurons depends on the number of neurons in the succeeding layer (e.g., the output layer in this example).
Generally speaking, the number of output neurons is significantly smaller than the number of intermediate neurons. 
Therefore we conduct the reduction in a backward manner from the last hidden layer to the first hidden layer, and the experiments in \autoref{sec:experiment} show that a major proportion of the neurons could be deleted, which boosts verification efficiency and therefore improve precision within a given execution timeout. 

\section{Stable ReLU Neurons Reduction}
\label{sec:stableandreduce}

\subsection{Stable ReLU neurons}
Given a ReLU layer $X$ in a neural network and an input $\vec{a}$, the layer $X$ has exactly one preceding layer $Y$ (i.e. $in(X)=\{Y\}$) and the {\em pre-activation} of the $k^{th}$ ReLU neuron $x$ in $X$ is the output $y(\vec{a})$ of the $k^{th}$ neuron $y$ in $Y$.
For simplicity, we use $\hat x(\vec{a})=y(\vec{a})$ to denote the pre-activation of $x$. 

\definition{
A ReLU neuron $x$ in a neural network is {\em deactivated} w.r.t. (with respect to) an input space $I$ if $\hat x(\vec{a})\le 0$ for all inputs $\vec{a}\in I$,
and $x$ is {\em activated} w.r.t. the input space $I$ if $\hat x(\vec{a})\ge 0$ for all $\vec{a}\in I$.
Then $x$ is {\em stable} w.r.t. $I$ if $x$ is deactivated or activated w.r.t. $I$.
}

It is NP-complete to check whether the output of a neural network is greater than or equal to 0 \cite{DBLP:journals/corr/KatzBDJK17}.
In addition, we can add an additional ReLU layer behind the output layer of the neural network where the output of the original neural network becomes the pre-activation of ReLU neurons, therefore, it is straightforward that checking the stability of ReLU neurons w.r.t. an input space is NP-hard.

\theorem{
It is NP-hard to check whether a ReLU neuron is stable w.r.t. an input space $I$.
}

\begin{table}[!hb]
\centering
\caption{Bound propagation methods}
\label{tab:BoundPropMethods}
\vspace{-0.5em}
  \addtolength{\tabcolsep}{-0.5em}
\begin{tabular}{|c|c|c|c|}
\hline
Methods for Stability Detection & Other Methods\\
\hline
\hline
Interval \cite{gowal2018effectiveness}, DeepZ/Symbolic \cite{singh2018fast,weng2018towards,wang2018efficient}&$\beta$-Crown \cite{DBLP:conf/nips/WangZXLJHK21}\\
CROWN\cite{DBLP:conf/nips/ZhangWCHD18}, FCrown \cite{lyu2020fastened}, $\alpha$-Crown \cite{xu2021fast} &    GCP-Crown \cite{zhang2022general}\\
Deepoly\cite{DBLP:journals/pacmpl/SinghGPV19}, kPoly \cite{DBLP:conf/nips/SinghGPV19}, PRIMA \cite{DBLP:journals/pacmpl/MullerMSPV22}& ARENA \cite{DBLP:conf/vmcai/ZhongTK23}\\
RefineZono \cite{singh2019boosting},OptC2V \cite{tjandraatmadja2020convex}&  DeepSRGR \cite{DBLP:conf/tacas/YangLLHWSXZ21}\\
SDP-Relaxation \cite{raghunathan2018semidefinite,batten2021efficient,lan2022tight}&  \\
LP/Lagrangian Dual \cite{wong2018provable,dvijotham2018dual,bunel2020lagrangian}&  \\
\hline
\end{tabular}
\end{table}

Many  methods have been proposed to compute the lower and upper bounds of the pre-activation of ReLU neurons.
Usually, these methods use different constraints to tighten the pre-activation bounds, such as linear relaxations, split constraints, global cuts and output constraints.
For detecting the stability of ReLU neurons w.r.t. an input space, we only consider the methods which employ the linear relaxations of ReLU neurons alone, as the other constraints may filter out inputs from the input space, e.g. a ReLU neuron is compulsory to be stably deactivated despite the input space if the propagator uses a split constraint to enforce that the pre-activation input value is always non-positive.
%
%
We enumerate various bound propagation methods in \autoref{tab:BoundPropMethods}.
The methods using other constraints are not suitable for detecting the stability of intermediate neurons, such as $\beta$-Crown employs split constraints.

In our experiments, we use the GPU-based bound propagation method CROWN to detect stable ReLU neurons, which leads to a reasonably significant reduction with a small time cost, as displayed in \autoref{tab:reduceSizeRes}.
\vspace{-0.3em}
\subsection{ReLU layer reduction}
As illustrated in the example given in \autoref{sec:REDoverview}, after computation of concrete neuron bounds, we detect and handle those stable ReLU neurons in the following ways:
\begin{itemize}
    \item For a stably deactivated ReLU neuron whose input value is always non-positive, it is always evaluated as $0$ and thereby will be directly deleted from the network as it has no effect on the actual computation;
    \item For a stably activated ReLU neurons (the input values of which are always non-negative) in the same layer, we reconstruct this set of stably activated neurons into \emph{a smaller} set of stably activated neurons as we reduce $x_4, x_5, x_6$ into $m_1, m_2$ in \autoref{sec:REDoverview}.
\end{itemize}
\textbf{Reconstruction of Stably Activated Neurons. } \autoref{fig:removalNet} illustrates that the deletion of stably activated neurons requires creating new connections between the preceding and succeeding neurons of the deleted neurons.
We follow the convention that every intermediate ReLU layer only directly connects to one preceding layer and one succeeding layer, which conducts linear computation (and we defer the details of how to simplify a complicated network with multiple preceding/succeeding connections into such simpler architecture in \autoref{sec:simplyMethod}).

\begin{figure}[!ht]
\scriptsize
\centering
\begin{tikzpicture}[
    layer/.style={rectangle, draw = black, thin,
      minimum width=4mm, minimum height=2cm}
    ]
  \node[layer, black] (v){};
  \node[black, above = 3mm of v](){Layer $V$};
  \node[black, below = 1.5mm of v](){$q$ neurons};
  \node[layer, black, right = 15mm of v] (x){};
  \node[black, above = 3.5mm of x](){Affine $X$};
  \node[black, above = 0mm of x](){$B_1$};
  \node[black, below = 1.5mm of x](){$m$ neurons};
  \node[layer, black, right = 15mm of x] (y){};
  \node[black, above = 3mm of y](){ReLU $Y$};
  \node[black, below = 1.5mm of y](){$m$ neurons};
  
  \node[layer, black, right = 15mm of y] (z){};
  \node[black, above = 0mm of z](){$B_2$};
  \node[black, above = 3mm of z](){Affine $Z$};
  \node[black, below = 1.5mm of z](){$n$ neurons};

  \node (C0) at ($(v)!0.5!(x)$) {};
  \node[black, above = -1 mm of C0] (q1) {$M_{1}$};
  \node (C1) at ($(x)!0.5!(y)$) {};
  \node[black, above = -1 mm of C1] () {$ReLU()$};
  \node (C2) at ($(y)!0.5!(z)$) {};
  \node[black, above = -1 mm of C2] (q2) {$M_{2}$};
  \draw [->,black, thick](v) -- (x);
  \draw [->,black, thick](x) -- (y);
  \draw [->,black, thick](y) -- (z);
\end{tikzpicture}
\vspace{-0.5em}
\caption{Layer $X$ and $Y$ with pending reduction, together with its preceding layer $V$ and succeeding affine layer $Z$.}
\label{fig:exampleblock}
\end{figure}
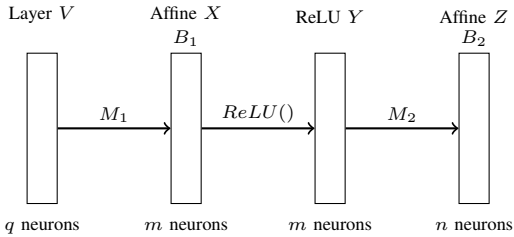

An example of a ReLU layer pending reduction is shown in \autoref{fig:exampleblock}, where $M_{1}$ indicates the linear connection between layer $V$ and $X$; $M_{2}$ indicates the connection between layer $Y$ and $Z$.
Biases are recorded in $B_{1}$ and $B_{2}$ respectively.
Suppose that the uppermost $k$ neurons in layer $Y$ are stably activated, and we delete them together with their inputs in layer $X$ from \autoref{fig:exampleblock}.
After deletion, we need to generate a new connection between layers $Z$ and $V$.
As stably activated ReLU neurons behave as identity functions, the new connection matrix between layer $Z$ and $V$ can be computed from existing connection matrices $M_{1}$ (size $m \times q$) and $M_{2}$ (matrix with size $n \times m$). 
Assume that $M[0:k, :]$ indicates that we slice the matrix to contain only the first $k$ rows; and $M[:, 0:k]$ means we only take the leftmost $k$ columns of the matrix, we define a matrix $M'_{VZ}$ with size $n \times q$ that is computed as:
\vspace{-0.3em}
\begin{equation}
    M'_{VZ} = M_{2}[:, 0:k] \cdot M_{1}[0:k, :] \label{equ:collapsetoLinear}
\end{equation}
\vspace{-0.5em}

\begin{figure}[!ht]
\scriptsize
\centering
\begin{tikzpicture}[
    layer/.style={rectangle, draw = black, thin,
      minimum width=4mm, minimum height=2cm},
    del_layer/.style={rectangle, draw = black, thin,
      minimum width=4mm, minimum height=0.8cm},
    merge_layer/.style={rectangle, draw = black, thin, densely dashed,
      minimum width=4mm, minimum height=0.5cm}
    ]
  \node[layer, black] (v){};
  \node[black, above = 3mm of v](){Layer $V$};
  \node[black, below = 1.5mm of v](){$q$ neurons};
  \node[del_layer, black, below right = -9mm and 20mm of v] (x){};
  \node[black, above = 14mm of x](){Affine $X$};
  \node[merge_layer, blue, above = 4mm of x] (m1){$M$};
  \node[blue, above = 0.5mm of m1](){$B$};
  \node[black, above = -0.5mm of x](){$B_{1}[k:m]$};
  \node[black, below = 1.5mm of x](){$m-k+n$ neurons};
  \node[del_layer, black, right = 18mm of x] (y){};
  \node[black, above = 14mm of y](){ReLU $Y$};
  \node[merge_layer, blue, above = 4mm of y] (m2){$M'$};
  \node[black, below = 1.5mm of y](){$m-k+n$ neurons};
  
  \node[layer, black, right = 65mm of v] (z){};
  \node[blue, above = -0mm of z](){$B' + B_2 - B$};
  \node[black, above = 3mm of z](){Affine $Z$};
  \node[black, below = 1.5mm of z](){$n$ neurons};

  \node (C0) at ($(v)!0.5!(x)$) {};
  \node[black, above = -3 mm of C0] () {$M_{1}[k:m, :]$};
  \node (C1) at ($(x)!0.5!(y)$) {};
  \node[black, above = -1 mm of C1] () {$ReLU()$};

  \node (C2) at ($(y)!0.5!(z)$) {};
  \node[black, above = -3 mm of C2] () {$M_{2}[:, k:m]$};
  \node (C3) at ($(0.25,0.9)!0.5!(8.05,0.9)$) {};
    \node (C4) at ($(m1)!0.5!(m2)$) {};
  \node[black, above = -1 mm of C4] () {$ReLU()$};
  \node (C5) at ($(0.2, 0.55)!0.5!(m1)$) {};
  \node[black, above = -1 mm of C5] () {$M'_{VX}$};
    \node (C6) at ($(m2)!0.5!(6.7, 0.55)$) {};
  \node[black, above = -1 mm of C6] () {$M_I$};
  \draw [->,black, thick](0.2,-0.5) -- (x);
  \draw [->,black, thick](0.2, 0.55) -- (m1);
  \draw [->,black, thick](x) -- (y);
  \draw [->,black, thick](m1) -- (m2);
  \draw [->,black, thick](y) -- (6.7,-0.5);
   \draw [->,black, thick] (m2) -- (6.7, 0.55);
\end{tikzpicture}
\caption{The block after reduction of stably activated neurons. $M_{1}[k:m, :]$ contains the last $m-k$ rows of $M_{1}$, while $M_{2}[:, k:m]$ takes the rightmost $m-k$ columns of $M_{2}$. 
$B'$ is computed as $M_{2}[:, 0:k]\cdot B_{1}[0:k]$.
The \emph{newly constructed} neurons are dashed and colored in blue.}
\label{fig:afteractdeletion}
\end{figure}
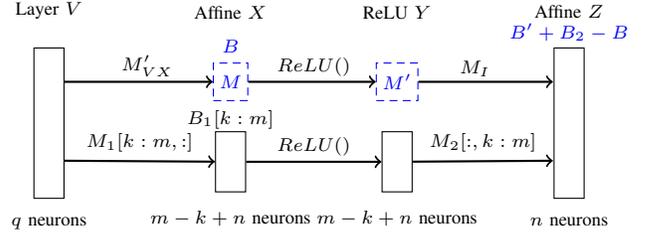

We consider this new connection $M'_{VZ}$ with size $n \times q$ as:
\vspace{-0.3em}
\begin{equation}
    M'_{VZ} = M_I \cdot M'_{VX} \label{equ:svd} ,
\end{equation}
where $M'_{VX}$ {\em equals to} $M'_{VZ}$ and functions as the affine connection between layers $V$ and \emph{newly constructed} neurons in layer $X$; $M_I$ denotes an $n \times n$ identity matrix and is the affine connection between layer $Z$ and the \emph{newly constructed} neurons in layer $Y$, as shown in \autoref{fig:afteractdeletion}. 
Here, the additional weight matrix between layers $V$ and $Z$ is actually computed as  $M'_{VZ} = M_I \cdot ReLU() \cdot M'_{VX} $.
For \autoref{equ:svd} to hold, we need to make sure that the ReLU function between $M$ and $M'$ becomes an identity, which means $M$ must be non-negative and $M'$ is stably activated.
So we will compute the concrete bounds of $M$ and add an additional bias $B$ to enforce it as non-negative as we did for neuron $m_1$ in \autoref{sec:REDoverview}.
This additional bias will be canceled out at layer $Z$ with $-B$ offset.

Note that we conduct this reduction in a backward manner from the last hidden layer (whose succeeding layer is the output layer that usually consists of a very small number of neurons, e.g. 10) to the first hidden layer.
Therefore, upon reduction of layers $X$ and $Y$, layer $Z$ has already been reduced and contains a small number $n$ of neurons. 
In the end, the $k$ stably activated neurons will be reduced into $n$ stably activated neurons and we obtain a smaller-sized affine layer with $m-k+n$ neurons, where $k$ is usually much bigger than $n$. 
Therefore, we are able to observe a significant size reduction as shown in \autoref{tab:reduceSizeRes}.


\lemma{
\label{lemma:reduceEuqal}
%
%
The reduction process preserves the input-output equivalence of the original network.
That is, for any input $\vec{a} \in I$, $\mathcal{F}(\vec{a}) \equiv \mathcal{F'}(\vec{a})$ where $\mathcal{F}$ is the original network and $\mathcal{F'}$ is the reduced one.
}
\vspace{-0.3em}
\begin{proof}
The reduction process operates on ReLU neurons that are stable w.r.t. the input space $I$. 
Specifically,
(i) Stably {\em deactivated} ReLU neurons are always evaluated as $0$ and can be deleted directly as they have no effect on the subsequent computation;
(ii) Stably {\em activated} ReLU neurons are reconstructed in a way that their functionality are preserved before (\autoref{fig:exampleblock}) and after (\autoref{fig:afteractdeletion}) reconstruction.

%
For any $\vec{a} \in I$, $V(\vec{a})$ is the output of Layer $V$ and the output of Layer $Z$ is computed as 
$Z(\vec{a})=M_{2}\cdot ReLU(M_{1}\cdot V(\vec{a}) + B_{1}) + B_{2}$ in \autoref{fig:exampleblock}.
%
we decompose $Z(\vec{a})-B_2$
as 
\begin{align}
  M_{2}[:, 0:k]\cdot ReLU(M_{1}[0:k,:]\cdot V(\vec{a}) + B_{1}[0:k])\label{equ:stableterms}  \\
 + M_{2}[:, k:m]\cdot ReLU(M_{1}[k:m,:]\cdot V(\vec{a})+ B_{1}[k:m])
\end{align}

Without loss of generality , we assume the uppermost $k$ neurons in layer
$Y$ are stably activated.
Formula \ref{equ:stableterms} thus simplifies to $M_{2}[:, 0:k]\cdot M_{1}[0:k,:] \cdot V(\vec{a}) + M_{2}[:, 0:k]\cdot B_{1}[0:k]  = M_I \cdot M'_{VX} \cdot V(\vec{a}) + B'$,
%
where $M'_{VX} = M_{2}[:, 0:k]\cdot M_{1}[0:k,:]$, $B'=M_{2}[:, 0:k]\cdot B_{1}[0:k]$ and $M_I$ is an identity matrix.
Furthermore, we compute an additional bias $B$ to ensure that $M'_{VX} \cdot V(\vec{a}) + B \geq 0$ for all $\vec{a} \in I$. 
Thus Formula \ref{equ:stableterms} finally simplifies to:
\begin{equation}
M_I \cdot ReLU( M'_{VX} \cdot V(\vec{a}) + B) -  B +B'
\label{equ:afterreconstruction}
\end{equation}
%
Based on Formula \ref{equ:afterreconstruction}, we obtain $Z(\vec{a}) = M_I \cdot ReLU(M'_{VX} \cdot V(\vec{a}) + B) + M_{2}[:, k:m]\cdot ReLU(M_{1}[k:m,:]\cdot V(\vec{a})+ B_{1}[k:m]) +B' + B_2 - B$, which equals to the computation conducted in \autoref{fig:afteractdeletion}. 
Thus,
the network preserves input-output equivalence after reduction.
\end{proof}

\begin{figure}[t]
\scriptsize
\centering
\begin{tikzpicture}[->, >=stealth']
\node[state] (DNN) 
{
\begin{tabular}{l}
\textbf{Input NN:}\\
ReLU, MaxPooling,\\
Conv, GeMM, Add,\\
Sub, Concat, Reshape,\\
Flatten, MatMul, \\
BatchNormalization, $\cdots$\\
\end{tabular}
};

\node[state, node distance = 5.5cm, right of= DNN] (INN) 
{
\begin{tabular}{l}
\textbf{Intermediate NN:}\\
ReLU,\\ 
SumLinear\\
\end{tabular}
};

\node[state, node distance = 2.5cm, below of= INN, xshift = 0cm] (SDNN) 
{
\begin{tabular}{l}
\textbf{Simple NN:}\\
ReLU, Linear\\
\end{tabular}
};

\node[state, node distance = 3.2cm, left of= SDNN, xshift = 0cm] (REDNet) 
{
\begin{tabular}{l}
\textbf{REDNet:}\\
ReLU, GeMM\\
\end{tabular}
};

\node[state, node distance = 3cm, left of= REDNet] (V) 
{
\begin{tabular}{l}
\textbf{Verifiers:}\\
$\alpha$,$\beta$-crown,\\
PRIMA,\\
VeriNet,\\
$\cdots$\\
\end{tabular}
};

\path[->]
(DNN) edge[] node[yshift=0.3cm] {\textbf{encode}} (INN)
(INN) edge[bend left] node[yshift=0cm] {\textbf{transform}} (SDNN)
(SDNN) edge[] node[yshift=0.3cm] {\textbf{reduce}} (REDNet)
(REDNet) edge[] node[yshift=0.3cm] {\textbf{verify}} (V);

\end{tikzpicture}
\vspace{-0.5em}
\caption{The procedure of neural network reduction. The encoding session is described in \autoref{sec:encodedes}; the transformation is discussed in \autoref{sec:transformdis}; and the reduction part is explained in \autoref{sec:stableandreduce}.}
\label{fig:framework}
\end{figure}
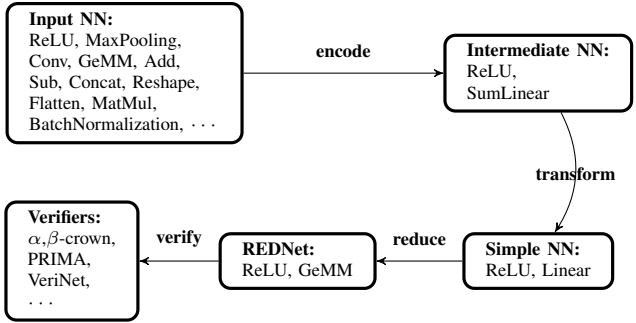







\vspace{-1em}
\section{Neural Network Simplification}
\label{sec:simplyMethod}
In \autoref{sec:stableandreduce}, we describe how reduction is conducted on a sequential neural network, where each intermediate layer only connects to one preceding Linear layer and one succeeding Linear layer.
In this paper, a Linear layer refers to a layer whose output is computed via linear computation.
Nonetheless, there exist many complicated network architectures (e.g., residual networks) that are not sequential.
In order to handle a wider range of neural networks, we propose a neural network simplification process to transform complex network architectures into simplified sequential neural networks and then conduct reduction on the simplified network.

We now introduce how to transform a complex ReLU-based neural network into a sequential neural network consisting of Linear and ReLU layers so that stable ReLU neurons can be reduced.
Note that we only consider the neural network layers that can be encoded as Linear and ReLU layers; further discussion about this can be found in \autoref{sec:discussion}.

The network simplification process involves two main steps (shown in Figure \ref{fig:framework}): (i) Encode various layers as SumLinear blocks and ReLU layers (we defer the definition of SumLinear block to \autoref{sec:encodedes});
(ii) Transform SumLinear blocks into Linear layers.
Here, Linear layers refer to layers that conduct linear computation. 
%
\subsection{Encode various layers into SumLinear blocks}
\label{sec:encodedes}
A {\em SumLinear block} is a combination of a set of Linear layers and a Sum layer such that the Linear layers are preceding layers of the Sum layer, where the output of the Sum layer is the element-wise sum of its inputs. 
%
%
The output of the SumLinear block is equal to the element-wise sum of the outputs of the Linear layers, and the preceding layers of the SumLinear block include the preceding layers of all the Linear layers.
%
Any Linear layer can be encoded as a SumLinear block by adding a Sum layer behind the Linear layer.
A main difference between them is that the SumLinear block can have more than 1 preceding layer. 

Many neural network layers can be directly transformed into SumLinear blocks, such as Conv, GeMM, Add, Sub, Concat, Reshape, Split, Squeeze, Unsqueeze, $\cdots$, Flatten, MatMul, and BatchNormalization layers used in ONNX models.\footnote{In general, Maxpooling can be encoded as Conv and ReLU layers with existing tools such as DNNV \cite{DNNVSystem}. Note that $max(x,y)=ReLU(x-y)+y$.}
Note that the Linear layer only has one preceding layer, while the Add and Concat layers can have more than one preceding layer; hence, they cannot be directly encoded as a Linear layer (this motivates the introduction of SumLinear blocks). 

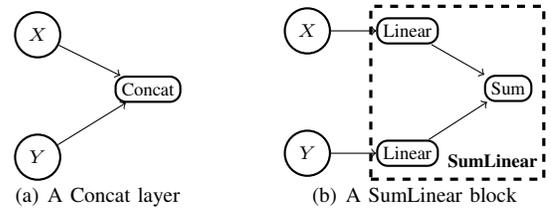
\begin{figure}[!ht]
\scriptsize
\centering
\subfigure[A Concat layer]{
\begin{tikzpicture}[
    black_node/.style={circle, draw = black, thick,
      minimum size = 6mm, inner sep=1pt},
    black_rectangle/.style={rectangle, 
        rounded corners,
        draw=black, thick,
        text centered},
    black_dash_rectangle/.style={rectangle, draw = black, thin, dashed}
    ]
  \node[black, black_node] (x1){$X$};
  \node[black, black_node, below = 10mm of x1](x2){$Y$};
  \node[black, black_rectangle, above right = 5mm and 8mm of x2] (x3){Concat};
  \draw [->,black,thin](x1) -- (x3);
  \draw [->,black,thin](x2) -- (x3);
\end{tikzpicture}
}
\hspace{0.4in}
\subfigure[A SumLinear block]{
\begin{tikzpicture}[
    black_node/.style={circle, draw = black, thick,
      minimum size = 6mm, inner sep=1pt},
    black_rectangle/.style={rectangle, 
        rounded corners,
        draw=black, thick,
        text centered},
    black_dash_rectangle/.style={rectangle, draw = black, very thick, dashed, minimum size = 23mm}
    ]
  \node[black, black_node] (x1){$X$};
  \node[black, black_node, below = 10mm of x1](x2){$Y$};
  \node[black, black_rectangle, right = 6mm of x1] (x3){Linear};
  \node[black, black_rectangle, right = 6mm of x2](x4){Linear};
  
  \node[black, black_rectangle, above right = 5mm and 6mm of x4] (x5){Sum};
  \node[black, black_dash_rectangle, below right = 5mm and 5mm of x1, yshift = 10.7mm, xshift=1mm] (x6){};
  \node[black, right = 0.3mm of x4, yshift=-1mm] () {\textbf{SumLinear}};
  \draw [->,black,thin](x1) -- (x3);
  \draw [->,black,thin](x2) -- (x4);
  \draw [->,black,thin](x3) -- (x5);
  \draw [->,black,thin](x4) -- (x5);
\end{tikzpicture}
}
\vspace{-0.5em}
\caption{Encode a Concat layer into a SumLinear block.}
\label{fig:concat}
\end{figure}

Figure \ref{fig:concat} shows a SumLinear block encoding a Concat layer with 2 precedessors $X$ and $Y$.
The biases of the two Linear layers are zero and the concatenation of their weights is an identity matrix. 
Thus, each neuron of the layers $X,Y$ is mapped to a neuron of the Sum layer.
Assume $|X| = |Y| = 1$.
Their weights are represented by matrices:
$\begin{bmatrix}
1\\
0\\
\end{bmatrix}$
and 
$\begin{bmatrix}
0\\
1\\
\end{bmatrix}$ 
which can be concatenated into an identity matrix $\begin
{bmatrix}
1&0\\
0&1\\
\end{bmatrix}$.

\begin{figure}[!ht]
\scriptsize
\centering
\subfigure[An Add layer]{
\begin{tikzpicture}[
    black_node/.style={circle, draw = black, thick,
      minimum size = 6mm, inner sep=1pt},
    black_rectangle/.style={rectangle, 
        rounded corners,
        draw=black, thick,
        text centered},
    black_dash_rectangle/.style={rectangle, draw = black, thin, dashed}
    ]
  \node[black, black_node] (x1){$X$};
  \node[black, black_node, below = 10mm of x1](x2){$Y$};
  \node[black, black_rectangle, above right = 5mm and 8mm of x2] (x3){Add};
  \draw [->,black,thin](x1) -- (x3);
  \draw [->,black,thin](x2) -- (x3);
\end{tikzpicture}
}
\hspace{0.4in}
\subfigure[A SumLinear block]{
\begin{tikzpicture}[
    black_node/.style={circle, draw = black, thick,
      minimum size = 6mm, inner sep=1pt},
    black_rectangle/.style={rectangle, 
        rounded corners,
        draw=black, thick,
        text centered},
    black_dash_rectangle/.style={rectangle, draw = black, very thick, dashed, minimum size = 23mm}
    ]
  \node[black, black_node] (x1){$X$};
  \node[black, black_node, below = 10mm of x1](x2){$Y$};
  \node[black, black_rectangle, right = 6mm of x1] (x3){Linear};
  \node[black, black_rectangle, right = 6mm of x2](x4){Linear};
  
  \node[black, black_rectangle, above right = 5mm and 6mm of x4] (x5){Sum};
  \node[black, black_dash_rectangle, below right = 5mm and 5mm of x1, yshift = 10.7mm, xshift=1mm] (x6){};
  \node[black, right = 0.3mm of x4, yshift=-1mm] () {\textbf{SumLinear}};
  \draw [->,black,thin](x1) -- (x3);
  \draw [->,black,thin](x2) -- (x4);
  \draw [->,black,thin](x3) -- (x5);
  \draw [->,black,thin](x4) -- (x5);
\end{tikzpicture}
}
\vspace{-0.5em}
\caption{Encode an Add layer into a SumLinear block.}
\label{fig:addblock}
\end{figure}
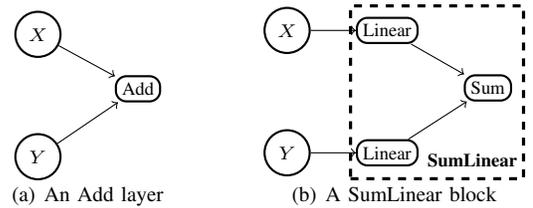

\vspace{1em}
In the same spirit, the Add layer could also be encoded as a SumLinear block as shown in \autoref{fig:addblock}.
Assume that $|X|$ and $|Y|$ are each equal to 2, the weights of the two Linear layers are represented by identity matrices
%
$\begin{bmatrix}
1&0\\
0&1\\
\end{bmatrix}$.
%
%
%



\subsection{Transform SumLinear blocks into Linear Layers}
\label{sec:transformdis}
In this subsection, we show how to encode SumLinear blocks as Linear layers.
Firstly, we need to transform SumLinear blocks into normalized SumLinear blocks. 
To this end, a SumLinear block $L$ is {\em normalized} if it does not have any Linear layer of which the preceding layer is a SumLinear block (i.e. $in(L)$ does not have SumLinear blocks), and each of the Linear layers in $L$ has different preceding layers.
For example, the SumLinear given in \autoref{fig:concat} is normalized if its preceding layers $X$ and $Y$ are not SumLinear blocks.

{\bf SumLinear Block Normalization}. 
If a SumLinear block $L'$ includes a Linear layer $L'_j$ with a weight $M'_j$ and a bias $B'_j$ such that the preceding layer of $L'_{j}$ is another SumLinear block $L''$ including $k$ Linear layers $L''_1,\cdots L''_k$ with weights $M''_1,\cdots,M''_k$ and biases $B''_1,\cdots,B''_k$, then we can normalize $L'$ by replacing $L'_j$ with $k$ new Linear layers $L_1,\cdots L_k$ where for any $1\le i\le k$, the layer $L_i$ has the same preceding layer as that of $L''_i$, and the weight and bias of $L_i$ are computed as:
\begin{align}
M_i=M'_j\cdot M''_i\hspace{2.73cm}\\
B_i= 
\begin{cases}
B'_j+M'_j\cdot B''_i &\text{if $i=1$}\\
M'_j\cdot B''_i &\text{otherwise}\\
\end{cases}
\end{align}
%
During the normalization, if the succeeding layers of the block $L''$ become empty, then $L''$ is directly removed.

In addition, if two Linear layers $L_a,L_b$ in a SumLinear block have the same preceding layer, then in normalization, we can replace them by one new Linear layer $L_c$ such that $L_c$ has the same preceding layer as them and the weight (and bias) of $L_c$ is the sum of the weights (and biases) of $L_a,L_b$.

\lemma{
\label{lemma:normalize}
SumLinear block normalization does not change the functionality of a neural network. 
}
\vspace{-0.5em}
\begin{proof}[Proof]
Let $\vec{a}''_i$ be any input of a Linear layer $L''_i$ in the block $L''$ where $L''$ is the preceding layer of $L'_j$.
Thus, the input of $L'_j$ (called  $\vec{a}'_j$) equals to $\sum_{i=1}^k (M''_i\cdot\vec{a}''_i+B''_i)$.
Then the output of $L'_j$ is 
$B'_j+\sum_{i=1}^k (M'_j\cdot M''_i\cdot\vec{a}''_i+M'_j\cdot B''_i)$
which is equal to the sum of the outputs of the layers $L_1,\cdots L_k$.
Therefore, replacing $L'_j$ with $L_1,\cdots L_k$ does not change the output of the SumLinear block $L'$.

If the succeeding layers of $L''$ become empty, then removing $L''$ does not affect the outputs of other layers and the network.

In addition, the sum of the outputs of the two linear layers $L_a,L_b$ in a SumLinear Block with the same preceding layer is equal to the output of the new layer $L_c$, thus, the output of the block does not change after replacing $L_a,L_b$ with $L_c$.

So SumLinear block normalization does not change the functionality of a neural network.
\end{proof}

We next show how to encode normalized SumLinear blocks as Linear layers. 


{\bf Linear Layer Construction}. 
First, we say that a ReLU layer $L_i$ is {\em blocked} by a SumLinear block $L$ if $L$ is the only succeeding layer of $L_i$. 
Then, we use $\mathcal{R}_L$ to denote the set of ReLU layers blocked by the SumLinear block $L$. Let $\mathcal{P}_L$ include other preceding layers of $L$ which are not in $\mathcal{R}_L$.
If $L$ is normalized, then $L$ and the set of ReLU layers in $\mathcal{R}_L$ 
can be replaced by a Linear layer $L^l$, a ReLU layer $L^r$ and a new SumLinear block $L^s$ such that

\begin{itemize}
\item the weight $M^l$ (the bias $B^l$) of the linear layer $L^l$ is a concatenation (the sum) of the weights (the bias) of the Linear layers in $L$ and the preceding layer of $L^l$ is $L^r$ and $L^l$ has the same succeeding layers as $L$;
\item the SumLinear block $L^s$ encodes a concatenation of layers in $\mathcal{P}_L$ and the preceding layers of layers in $\mathcal{R}_L$;
%
\item 
$L^s$ is the preceding layer of $L^r$.
\end{itemize}
Additionally, in order to make sure that the outputs of the layers in $\mathcal{P}_L$ can pass through the ReLU layer $L^r$, the neurons in $L^r$ which connect to the layers in $P_L$ are enforced as activated neurons by adding an additional bias $B$ to a Linear layer in $L^s$ and minus $M^l\cdot B$ from the bias of $L^l$.


\lemma{
\label{lemma:construction}
Linear layer construction does not change the functionality of a neural network.
}
\vspace{-0.5em}
\begin{proof}
The pre-activation of $L^r$ is the output of $L^s$ that equals to $B$ plus the concatenation of the outputs of layers in $\mathcal{P}_L$ and the pre-activation of Layers in $\mathcal{R}_L$.
This ensures that the output of $L^r$ equals to $B$ plus the concatenation (call it $\Vec{a}$) of the outputs of layers in $\mathcal{P}_L$ and $\mathcal{R}_L$.
Next, the output of $L^l$ equals to $M^l\cdot (B+\Vec{a})+B^l-M^l\cdot B=M^l\cdot\Vec{a}+B^l$ which is equal to the output of original layer $L$.

In addition, $L$ is the only succeeding layer of layers in $\mathcal{R}_L$, so replacing $\mathcal{R}_L$, $L$ with the layers $L^s$, $L^r$, $L^l$ does not change the functionality of the neural network.
\end{proof}


\begin{algorithm}[b]
\caption{Neural Network Simplification} \label{code:nns}
\KwIn{A neural network $(\mathcal{V},\mathcal{E})$}
\KwOut{A sequential neural network}
$\mathcal{V},\mathcal{E}\leftarrow Initialization(\mathcal{V},\mathcal{E})$\; \label{code:line:1}
\While{$(\mathcal{V},\mathcal{E})$ has SumLinear blocks}{
Let $L$ be the last SumLinear block in $(\mathcal{V},\mathcal{E})$\;\label{code:line:3}
$\mathcal{V},\mathcal{E},L\leftarrow Normalization(\mathcal{V},\mathcal{E},L)$\;\label{code:line:4}
\If{$|in(L)|>1$ }{\label{code:line:5} 
$\mathcal{V},\mathcal{E}\leftarrow LinearLayerConstruction(\mathcal{V},\mathcal{E},L)$\;\label{code:line:6}
}
\Else{
$\mathcal{V},\mathcal{E}\leftarrow Linearization(\mathcal{V},\mathcal{E},L)$\; \label{code:line:8}
}
}
\Return $(\mathcal{V},\mathcal{E})$\;
\end{algorithm}

{\bf Network simplification.} We use \autoref{code:nns} to transform ReLU-based neural networks $(\mathcal{V},\mathcal{E})$ into a sequential neural network consisting of Linear and ReLU layers.
At \autoref{code:line:1}, the function $Initialization(\mathcal{V},\mathcal{E})$ encodes all layers in $\mathcal{V}$ as SumLinear blocks and ReLU layers.
Between \autoref{code:line:3} and \autoref{code:line:8}, the algorithm repeatedly selects the last SumLinear block $L$ in $\mathcal{V}$ and reconstructs $L$ into Linear layers, where a SumLinear block is {\em the last} block means there is not any path from it to another SumLinear block.
$(\mathcal{V},\mathcal{E})$ only has 1 output layer, and the Linear and ReLU layers only have 1 preceding layer, thus, there is only one last SumLinear block.

At \autoref{code:line:4}, the function $Normalization(\mathcal{V},\mathcal{E},L)$ is used to normalize the last SumLinear block $L$.
If the normalized $L$ has more than one preceding layer (i.e. $|in(L)|>1$), then the function $LinearLayerConstruction(\mathcal{V},\mathcal{E},L)$ is used to replace $L$ with the layers $L^l,L^r,L^s$ introduced in the Linear layer construction (at \autoref{code:line:6}),
otherwise the function $Linearization(\mathcal{V},\mathcal{E},L)$ is used to directly replace $L$ with the only Linear layer included in $L$ (at \autoref{code:line:8}).

In the rest of this subsection, we show that \autoref{code:line:6} in \autoref{code:nns} can only be visited at most $|\mathcal{V}|$ times, thus, the algorithm can terminate and generate an equivalent sequential neural network consisting of Linear and ReLU layers.

\lemma{\label{lemma:1} 
Assume $(\mathcal{V},\mathcal{E})$ is a neural network consisting of Linear, ReLU layers and SumLinear blocks and $L$ is the last SumLinear block in $\mathcal{V}$. 
If $|in(L)|>1$ and $in(L)$ does not have SumLinear blocks and Linear layers, then $\mathcal{R}_L$ is not empty.}
\vspace{-0.5em}
\begin{proof}
$(\mathcal{V}, \mathcal{E})$ only has one output layer and all layers behind $L$ have at most one preceding layer, thus, a path from any layer before $L$ to the output layer must pass $L$.

Let $L_i$ be the last ReLU layer in $in(L)$. If $L_i$ has a succeeding layer $L_j$ such that $L_j\neq L$, then $L$ must be in all paths from $L_j$ to the output layer, and there would be a ReLU layer in $in(L)$ included by a path from $L_j$ to $L$, which meant that $L_i$ was not the last layer in $in(L)$, a contradiction. Hence,
$L_i$ is in $\mathcal{R}_L$ and $\mathcal{R}_L\neq \emptyset$.
\end{proof}

Based on \autoref{lemma:1}, if $|in(L)|>1$, then $|\mathcal{R}_L|\ge 1$, thus, the number of ReLU layers before the last SumLinear block in $\mathcal{V}$ is decreased after replacing $L$ and the layers in $\mathcal{R}_L$ with the layers $L^l, L^r, L^s$ introduced in the Linear layer construction where $L^s$ becomes the last SumLinear block in $\mathcal{V}$.
So we can get that \autoref{code:nns} can terminate and generate a neural network consisting of Linear and ReLU layers.

\begin{theorem}
Algorithm \ref{code:nns} can terminate and generate a neural network consisting of Linear and ReLU layers.
\end{theorem}
\vspace{-0.5em}
\begin{proof}
From \autoref{lemma:1}, we know that \autoref{code:line:6} in \autoref{code:nns} reduces the number of ReLU neurons before the last SumLinear block in $\mathcal{V}$, therefore, it can only be visited at most $|\mathcal{V}|$ times.
Note that SumLinear block normalization (at \autoref{code:line:4}) does not affect ReLU layers and the layers behind $L$.

Then \autoref{code:line:8} can directly replace all SumLinear blocks having one preceding layer in $\mathcal{V}$ with Linear layers.
Therefore, \autoref{code:nns} can terminate and return a neural network consisting of Linear and ReLU layers.
\end{proof}

\begin{theorem}
\label{theorem:equivalence}
Our constructed REDNet is input-output equivalent to the original network given the input space $I$.
\end{theorem}
\vspace{-0.5em}
\begin{proof}
(Sketch.)
Our reduction technique contains two steps: (i) network simplification presented in \autoref{sec:simplyMethod}; (ii) stable ReLU neuron reduction described in \autoref{sec:stableandreduce}.
Each step is designed deliberately to preserve input-output equivalence.

{\em Simplification equivalence.} 
Function $Initialization(\mathcal{V},\mathcal{E})$  at \autoref{code:line:1} in \autoref{code:nns} encodes ONNX layers into a uniform network representation; such encoding preserves input-output equivalence. 
Then \autoref{lemma:normalize} and \autoref{lemma:construction} show that \autoref{code:line:4} and \autoref{code:line:6} do not change network functionality.
In addition, \autoref{code:line:8}, replacing a SumLinear Block with  the only Linear layer in the block, also does not change network output.
Therefore, \autoref{code:nns} can construct a sequential neural network that has the same functionality as the original neural network.

{\em Reduction equivalence.} The proof is given in \autoref{lemma:reduceEuqal}. 
\end{proof}

\subsection{Illustrative example of network simplification}
\begin{figure}[!ht]
\centering
\scriptsize
\subfigure[Before simplification]{
\begin{tikzpicture}[
    relu/.style={rectangle, draw = red, thick,
      minimum width=1cm, minimum height=0.5cm},
    conv/.style={rectangle, draw = blue, thick,
      minimum width=1cm, minimum height=0.5cm},
  add/.style={rectangle, draw = black, thick,
      minimum width=1cm, minimum height=0.5cm}
    ]
  \node[relu, red] (r1){$n_1$};
  \node[conv, blue, below left = 4.5mm and 0 mm of r1](c1){$n_2$};
  \node[relu, red, below = 4.5mm of c1](r2){$n_3$};
  \node[conv, blue, below = 4.5mm of r2](r4){$n_4$};
  \node[conv, blue, right = 8mm of r4](r3){$n_5$};
  \node[add, black, below right = 4.5mm and 0.5mm of r4](a1){$n_6$};
  \draw [->,black, thin](r1) -- (c1);
  \draw [->,black, thin](c1) -- (r2);
  \draw [->,black, thin](r2) -- (r4);
  \draw [->,black, thin](r4) -- (a1);
  \draw [black, thin](r1) -- (0.8, -0.8)  -- (r3);
  \draw [->, black, thin] (r3) -- (a1);
  \node () at (1.3, -2) {conv3};
  \node () at (-0.9, -0.4) {conv1};
  \node () at (-1.5, -1.5) {ReLU};
  \node () at (-1.5, -2.4) {conv2};
 \end{tikzpicture}
 \label{fig:blockinNodesandConnections}
}
\hspace{0.2in}
\subfigure[After simplification]{
   \begin{tikzpicture}[
    relu/.style={rectangle, draw = red, thick,
      minimum width=1cm, minimum height=0.5cm},
    conv/.style={rectangle, draw = blue, thick,
      minimum width=1cm, minimum height=0.5cm},
  add/.style={rectangle, draw = black, thick,
      minimum width=1cm, minimum height=0.5cm}
    ]
  \node[red] (origin){};
  \node[relu, red, right = 1mm of origin] (r1){$n_1$};
  \node[conv, blue, below = 8mm of r1](c1){$n_2 \oplus n_1'$};
  \node[relu, red, below = 8mm of c1](r2){$n_3 \oplus n_1''$};
  \node[conv, blue, below  = 8mm of r2](a1){$n_6$};
  \draw [->,black, thin](r1) -- (c1);
  \draw [->,black, thin](c1) -- (r2);
  \draw [->,black, thin](r2) -- (a1);
  \node () at (1.7, -0.65) {conv1$\oplus$identity};
  \node () at (1.3, -2) {ReLU};
  \node () at (1.6, -3.3) {conv2$\oplus$conv3};

\end{tikzpicture}
\label{fig:afterSimplification}
}
\vspace{-0.5em}
\caption{The simplification of a non-sequential block}
\label{fig:SimplyResiBlock}
\end{figure}
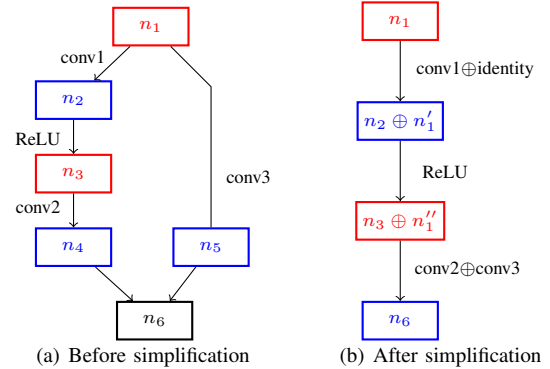

In this subsection, we use a simple network block to illustrate how to perform \autoref{code:nns} on a non-sequential structure (\autoref{fig:blockinNodesandConnections}) to get a sequential neural network consisting of Linear and ReLU layers (\autoref{fig:afterSimplification}).
In \autoref{fig:SimplyResiBlock}, each rectangular node (including $n_1, n_2, n_3, n_4$) represents a set of neurons whose values are derived from the preceding connected node(s) and the connections between them.
Note that red-colored rectangular nodes are ReLU nodes that represent the output neurons of the ReLU layer; blue nodes are convolutional nodes; the black node is an Add layer.
The connections between nodes are represented with directed edges, and the connected functions are displayed near the edges (e.g. conv1, ReLU).
Symbol $\oplus$ represents concatenation.

Firstly, we apply function $Initialization(\mathcal{V},\mathcal{E})$ at \autoref{code:line:1} to encode \autoref{fig:blockinNodesandConnections} as SumLinear blocks and ReLU layers, where the weights and biases of each Linear layer are displayed above the layer.
We name the two ReLU nodes $n_1, n_3$ as ReLU1, ReLU2 respectively.
\begin{figure}[!ht]
\scriptsize
\centering
\begin{tikzpicture}[
    black_rectangle/.style={rectangle, 
        rounded corners,
        draw=black, thick,
        text centered}
    ]
  \node[black, black_rectangle] (n1){ReLU1};
  \node[black, black_rectangle, right = 3mm of n1] (l1){Linear};
  \node[black, black_rectangle, right = 3mm of l1] (s1){Sum};
  \node[black, black_rectangle, right = 3mm of s1] (n3){ReLU2};
  \node[black, black_rectangle, right = 3mm of n3] (l2){Linear};
  \node[black, black_rectangle, right = 3mm of l2] (s2){Sum};
  \node[black, black_rectangle, right = 3mm of s2] (l3){Linear};
  \node[black, black_rectangle, below right = 5mm and 8mm of l3] (s3){Sum};
  \node[black, black_rectangle, below = 15mm of l2] (l4){Linear};
  \node[black, black_rectangle, below = 15mm of s2] (s4){Sum};
  \node[black, black_rectangle, below = 15mm of l3] (l5){Linear};
  \node[black, above = 1mm of l1](a1){conv1-weights};
  \node[black, above = 0mm of a1]{conv1-biases};
  \node[black, above = 1mm of l2](a2){conv2-weights};
  \node[black, above = 0mm of a2]{conv2-biases};
  \node[black, above = 1mm of l3](a3){identity matrix};
  \node[black, above = 0mm of a3]{bias=0};
  \node[black, above = 1.5mm of l5](a5){identity matrix};
  \node[black, above = 0mm of a5]{bias=0};
  \node[black, above = 1.5mm of l4](a4){conv3-weights};
  \node[black, above = 0mm of a4]{conv3-biases};
  \draw [->,black,thin](n1) -- (l1);
  \draw [->,black,thin](l1) -- (s1);
  \draw [->,black,thin](s1) -- (n3);
  \draw [->,black,thin](n3) -- (l2);
  \draw [->,black,thin](l2) -- (s2);
  \draw [->,black,thin](s2) -- (l3);
  \draw [->,black,thin](l3) -- (s3);
  \draw [->,black,thin](l4) -- (s4);
  \draw [->,black,thin](s4) -- (l5);
  \draw [->,black,thin](l5) -- (s3);
  \draw [black, thin](n1)  -- (0, -1.9);
  \draw [->, black, thin] (0, -1.9) -- (l4);
\end{tikzpicture}
\vspace{0.5em}
\caption{Network in \autoref{fig:blockinNodesandConnections} encoded with SumLinear blocks and ReLU layers}
\label{fig:netinSumLinear}
\end{figure}
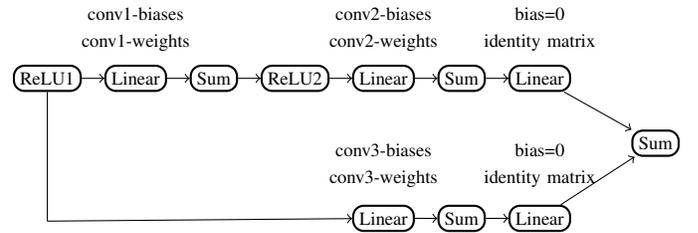

\begin{figure}[!ht]
\scriptsize
\centering
\begin{tikzpicture}[
    black_rectangle/.style={rectangle, 
        rounded corners,
        draw=black, thick,
        text centered}
    ]
  \node[black, black_rectangle] (n1){ReLU1};
  \node[black, black_rectangle, right = 5mm of n1] (l1){Linear};
  \node[black, black_rectangle, right = 5mm of l1] (s1){Sum};
  \node[black, black_rectangle, right = 5mm of s1] (n3){ReLU2};
  \node[black, black_rectangle, right = 5mm of n3] (l3){Linear};
  \node[black, black_rectangle, below right = 5mm and 12mm of l3] (s3){Sum};
  \node[black, black_rectangle, below = 15mm of l3] (l5){Linear};
  \node[black, above = 1mm of l1](a1){conv1-weights};
  \node[black, above = 0mm of a1]{conv1-biases};
  \node[black, above = 1mm of l3](a2){conv2-weights};
  \node[black, above = 0mm of a2]{conv2-biases};
  \node[black, above = 1mm of l5](a4){conv3-weights};
  \node[black, above = 0mm of a4]{conv3-biases};
  \draw [->,black,thin](n1) -- (l1);
  \draw [->,black,thin](l1) -- (s1);
  \draw [->,black,thin](s1) -- (n3);
  \draw [->,black,thin](n3) -- (l3);
  \draw [->,black,thin](l3) -- (s3);
  \draw [->,black,thin](l5) -- (s3);
  \draw [black, thin](n1)  -- (0, -1.9);
  \draw [->, black, thin] (0, -1.9) -- (l5);
\end{tikzpicture}
\caption{The network after the first normalization}
\label{fig:netstage1}
\end{figure}
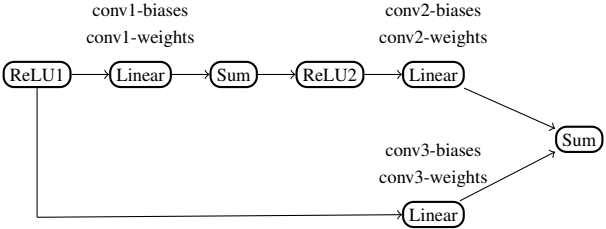

Then we take the last SumLinear block from \autoref{fig:netinSumLinear} and normalize this block (\autoref{fig:lastbeforeNormal}) and obtain the normalized block as in \autoref{fig:lastafterA1}.
The whole network is now updated as \autoref{fig:netstage1}.
\begin{figure}[!ht]
\scriptsize
\centering
\subfigure[The last block before normalization]{
\centering
\begin{tikzpicture}[
    black_rectangle/.style={rectangle, 
        rounded corners,
        draw=black, thick,
        text centered}
    ]
  \node[black, black_rectangle] (l2){Linear};
  \node[black, black_rectangle, right = 3mm of l2] (s2){Sum};
  \node[black, black_rectangle, right = 3mm of s2] (l3){Linear};
  \node[black, black_rectangle, below right = 5mm and 10mm of l3] (s3){Sum};
  \node[black, black_rectangle, below = 15mm of l2] (l4){Linear};
  \node[black, black_rectangle, below = 15mm of s2] (s4){Sum};
  \node[black, black_rectangle, below = 15mm of l3] (l5){Linear};
  \node[black, above = 1mm of l2](a2){conv2-weights};
  \node[black, above = 0mm of a2]{conv2-biases};
  \node[black, above = 1mm of l3](a3){identity matrix};
  \node[black, above = 0mm of a3]{bias=0};
  \node[black, above = 1mm of l5](a5){identity matrix};
  \node[black, above = 0mm of a5]{bias=0};
  \node[black, above = 1mm of l4](a4){conv3-weights};
  \node[black, above = 0mm of a4]{conv3-biases};
  \draw [->,black,thin](l2) -- (s2);
  \draw [->,black,thin](s2) -- (l3);
  \draw [->,black,thin](l3) -- (s3);
  \draw [->,black,thin](l4) -- (s4);
  \draw [->,black,thin](s4) -- (l5);
  \draw [->,black,thin](l5) -- (s3);
\end{tikzpicture}
\label{fig:lastbeforeNormal}
}
\subfigure[After normalization]{
\begin{tikzpicture}[
    black_rectangle/.style={rectangle, 
        rounded corners,
        draw=black, thick,
        text centered}
    ]
  \node[black, black_rectangle, right = 5mm of s2] (l3){Linear};
  \node[black, black_rectangle, below right = 5mm and 10mm of l3] (s3){Sum};
  \node[black, black_rectangle, below = 15mm of l3] (l5){Linear};
  \node[black, above = 1mm of l3](a2){conv2-weights};
  \node[black, above = 0mm of a2]{conv2-biases};
  \node[black, above = 1mm of l5](a4){conv3-weights};
  \node[black, above = 0mm of a4]{conv3-biases};
  \draw [->,black,thin](l3) -- (s3);
  \draw [->,black,thin](l5) -- (s3);
\end{tikzpicture}
\label{fig:lastafterA1}
}
\caption{Normalization of the last SumLinear block}
\end{figure}
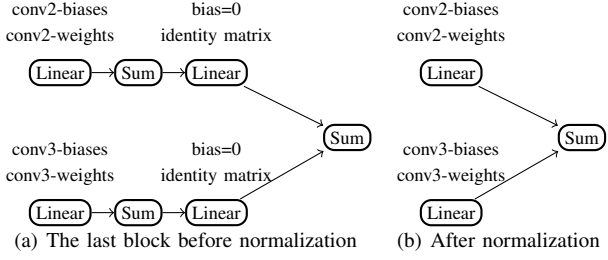

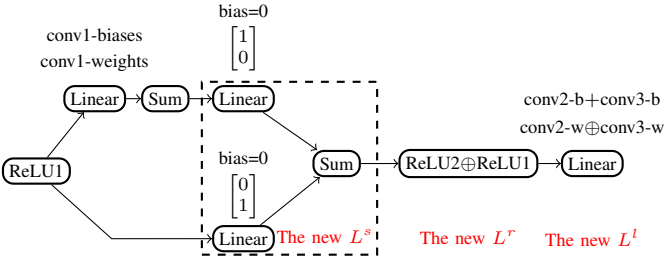
\begin{figure}[!ht]
\scriptsize
\centering
\begin{tikzpicture}[
    black_rectangle/.style={rectangle, 
        rounded corners,
        draw=black, thick,
        text centered},
    black_dash_rectangle/.style={rectangle, draw = black, thick, dashed, minimum width=2.33cm, minimum height=2.3cm}
    ]
  \node[black, black_rectangle] (n1){ReLU1};
  \node[black, black_rectangle, above right = 6mm and -1mm of n1] (l1){Linear};
  \node[black, black_rectangle, right = 2mm of l1] (s1){Sum};
  \node[black, black_rectangle, right = 3mm of s1] (l3){Linear};
  \node[black, black_rectangle, below right = 5mm and 5mm of l3] (s3){Sum};
  \node[black, black_rectangle, right = 5mm of s3] (new1){ReLU2$\oplus$ReLU1};
  \node[black, black_rectangle, right = 3mm of new1] (new2){Linear};
  \node[black, black_rectangle, below = 15mm of l3] (l5){Linear};
  \node[black, above = 1mm of l1](a1){conv1-weights};
  \node[black, above = 0mm of a1]{conv1-biases};
  \node[black, above = 1mm of l3](a2){$\begin{bmatrix} 1\\ 0\\\end{bmatrix}$};
  \node[black, above = -0.5mm of a2]{bias=0};
  \node[black, above = 0mm of l5](a4){$\begin{bmatrix} 0\\ 1\\\end{bmatrix}$};
  \node[black, above = -0.5mm of a4]{bias=0};
  \node[red, below = 6mm of new1](){The new $L^r$};
  \node[red, below = 6mm of new2](){The new $L^l$};
  \node[red, below left = 6mm and -9mm of s3](){The new $L^s$};
  \node[black, above = 1mm of new2](a3){conv2-w$\oplus$conv3-w};
  \node[black, above = 0mm of a3]{conv2-b$+$conv3-b};
  \node[black, black_dash_rectangle, below right = 5mm and 2mm of s1, yshift = 9.2mm, xshift=-0.5mm] (){};
  \draw [->,black,thin](n1) -- (l1);
  \draw [->,black,thin](l1) -- (s1);
  \draw [->,black,thin](s1) -- (l3);
  \draw [->,black,thin](l3) -- (s3);
  \draw [->,black,thin](l5) -- (s3);
  \draw [->,black,thin](s3) -- (new1);
  \draw [->,black,thin](new1) -- (new2);
  \draw [->,black,thin](n1) -- (1, -0.9) -- (l5);
\end{tikzpicture}
\caption{The network after the Linear layer construction. For simplicity to show the weight matrices, we assume that ReLU2 and ReLU1 all have one neuron; and ``-biases/-weights'' are abbreviated as ``-b/-w'' respectively.}
\label{fig:netstage3}
\end{figure}

At this step, we notice that ReLU layer ReLU2 is \textit{blocked} by the last SumLinear block, and ReLU1 is \textit{not blocked} as it has another path to a subsequent ReLU layer.
Therefore, we perform the Linear layer construction at \autoref{code:line:6} and obtain the network in \autoref{fig:netstage3}.

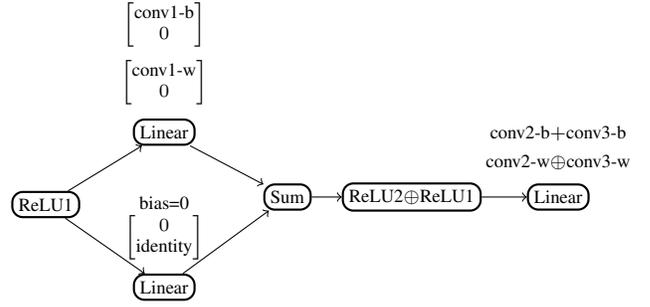
\begin{figure}[!ht]
\scriptsize
\centering
\begin{tikzpicture}[
    black_rectangle/.style={rectangle, 
        rounded corners,
        draw=black, thick,
        text centered},
    ]
  \node[black, black_rectangle] (n1){ReLU1};
  \node[black, black_rectangle, above right = 6mm and 7mm of n1] (l1){Linear};
  \node[black, black_rectangle, below right = 5mm and 9mm of l1] (s3){Sum};
  \node[black, black_rectangle, right = 4mm of s3] (new1){ReLU2$\oplus$ReLU1};
  \node[black, black_rectangle, right = 6mm of new1] (new2){Linear};
  \node[black, black_rectangle, below = 17mm of l1] (d1){Linear};
  \node[black, above = 1mm of l1](a2){$\begin{bmatrix} \text{conv1-w}\\ 0\\\end{bmatrix}$};
  \node[black, above = 1mm of d1](a4){$\begin{bmatrix} 0\\ \text{identity}\\\end{bmatrix}$};
 \node[black, above = 0mm of a2]{$\begin{bmatrix} \text{conv1-b}\\ 0\\\end{bmatrix}$};
  \node[black, above = -1mm of a4]{bias=0};
  \node[black, above = 1mm of new2](a3){conv2-w$\oplus$conv3-w};
  \node[black, above = 0mm of a3]{conv2-b$+$conv3-b};
  \draw [->,black,thin](n1) -- (l1);
  \draw [->,black,thin](l1) -- (s3);
  \draw [->,black,thin](n1) -- (d1);
  \draw [->,black,thin](d1) -- (s3);
  \draw [->,black,thin](s3) -- (new1);
  \draw [->,black,thin](new1) -- (new2);

\end{tikzpicture}
\caption{The network after the second normalization}
\label{fig:netstage4}
\end{figure}

Lastly, we take out the last SumLinear block in \autoref{fig:netstage3} and perform normalization to obtain \autoref{fig:netstage4}.
At \autoref{fig:netstage4}, the last SumLinear block includes two Linear layers having the same preceding layer ReLU1 (\autoref{fig:netstage4}), which requires further normalization.

\begin{figure}[!ht]
\scriptsize
\centering
\begin{tikzpicture}[
    black_rectangle/.style={rectangle, 
        rounded corners,
        draw=black, thick,
        text centered},
    ]
  \node[black, black_rectangle] (n1){ReLU1};
  \node[black, black_rectangle, right = 8mm of n1] (l1){Linear};
  \node[black, black_rectangle, right = 5mm of l1] (new1){ReLU2$\oplus$ReLU1};
  \node[black, black_rectangle, right = 9mm of new1] (new2){Linear};
  \node[black, above = 1mm of l1](a2){$\begin{bmatrix} \text{conv1-w}\\ \text{identity}\\\end{bmatrix}$};
 \node[black, above = 0mm of a2]{$\begin{bmatrix} \text{conv1-b}\\ 0\\\end{bmatrix}$};
  \node[black, above = 1mm of new2](a3){conv2-w$\oplus$conv3-w};
  \node[black, above = 0mm of a3]{conv2-b$+$conv3-b};
  \draw [->,black,thin](n1) -- (l1);
  \draw [->,black,thin](l1) -- (new1);
  \draw [->,black,thin](new1) -- (new2);
\end{tikzpicture}
\caption{The sequential network after the third normalization}
\label{fig:netstage5}
\end{figure}
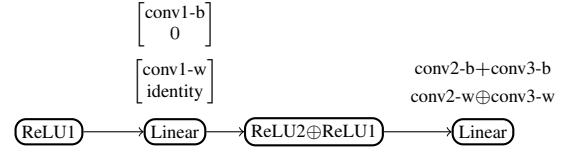

The final architecture of the network is given in \autoref{fig:netstage5}, where we have
$\begin{bmatrix} \text{conv1-w}\\ \text{identity}\\\end{bmatrix} = \text{conv1-w} \oplus \text{identity}$.
Now the oringal network has been simplified into a sequential one.

\section{Experiments}
\label{sec:experiment}
In this section, we present our experimental results of instantiation of network reduction technique on \crown{} \cite{AlphaBetaCrownSystem}, VeriNet \cite{VerinetSystem} and PRIMA \cite{DBLP:journals/pacmpl/MullerMSPV22} to show evidence that: given the \emph{same} verification problem, the \emph{same} verification algorithm runs \emph{faster} on the reduced network compared to the original network, which gives us confidence in the ability of our method as in enhancing the efficiency of existing verification methods.
Furthermore, the simple architecture in REDNet allows existing verification tools that only support limited network benchmarks to handle more networks.
\subsection{Experiment Setup}
The evaluation machine has two 2.40GHz Intel(R) Xeon(R) Silver 4210R CPUs with 384 GB of main memory and a NVIDIA RTX A5000 GPU. 

\textbf{Evaluation Benchmarks.} 
The evaluation datasets include MNIST \cite{DBLP:journals/spm/Deng12} and CIFAR10/CIFAR100 \cite{cifar10dataset}.  
MNIST dataset contains hand-written digits with 784 pixels, while CIFAR10/CIFAR100 includes colorful images with 3072 pixels.
We chose fully-connected, convolutional and residual networks with various sizes from two well-known benchmarks: the academic ERAN system \cite{ERANSystem} and VNNCOMP2021/2022 (International Verification of Neural Networks Competition) \cite{VNNCOMP21, VNNCOMP22}.
The number of activation layers (\#Layers), the number of ReLU neurons (\#Neurons), and the trained defense of each network are listed in \autoref{tab:netdetails}, where a trained defense refers to a defense method against adversarial samples to improve robustness.
Please note that ``Mixed" means mixed training, which combines adversarial training and certified defense training loss.
This could lead to an excellent balance between model clean accuracy and robustness, and is beneficial for obtaining higher
verified accuracy \cite{muller2022third}.

\begin{table}[!b]
\caption{Detailed information of the experimental networks}
  \centering
  \def\arraystretch{1.2}
   \addtolength{\tabcolsep}{-0.55em}
  \begin{tabular}{|l|c|c|c|c|c|c|}
    \hline
    \textbf{Network} &
    \textbf{Type} &
    \textbf{\#Layers} &
    \textbf{\#Neurons} &
    \textbf{Defense} &
    \textbf{\#Property}
    \\
    \hline
    M\_256x6 & fully-connected & 6 & 1,536 &  None & 30\\
    \hline
    M\_ConvMed & convolutional & 3 & 5,704 & None & 31\\
    \hline
     M\_ConvBig & convolutional & 6 & 48,064 &  DiffAI\cite{DBLP:conf/icml/MirmanGV18} & 29\\
    \hline
     M\_SkipNet & residual & 6 & 71,650 & DiffAI & 31\\
    \hline
     C\_8\_255Simp & convolutional  & 3 & 16,634 & None  & 30\\
    \hline
    C\_WideKW  & convolutional & 3 & 6,244 &  None  & 32 \\
    \hline
    C\_ConvBig & convolutional & 6 & 62,464 &  PGD\cite{DBLP:conf/iclr/MadryMSTV18} &  37 \\
    \hline
    C\_Resnet4b & residual  & 10 & 14,436 & None & 30\\
    \hline
    C\_ResnetA & residual  & 8 & 11,364 &  None & 32\\
    \hline
    C\_ResnetB & residual  & 8 & 11,364 &   None & 29 \\
    \hline
    C\_100\_Med& residual  & 10 & 55,460 &  Mixed & 24 \\
    \hline
    C\_100\_Large& residual  & 10 & 286,820 &  Mixed  & 24 \\
    \hline
  \end{tabular}
  \label{tab:netdetails}
\end{table}

\textbf{Verification Properties.} 
We conduct robustness analysis, where we determine if the classification result of a neural network – given a set of slightly perturbed images derived from the original image (input specification) – remains the same as the ground truth label obtained from the original unperturbed image (output specification).
The set of images is defined by a user-specified parameter $\epsilon$, which perturbs each pixel $p_i$ to take an intensity interval $[p_i - \epsilon, p_i + \epsilon]$.
Therefore, the input space $I$ is $\varprod_{i=1}^n [p_i-\epsilon, p_i+\epsilon]$.
In our experiment, we acquire the verification properties from the provided vnnlib files \cite{VNNLIBweb} that record the input and output specification or via a self-specified $\epsilon$.
We aim to speed up the analysis process for those {\em properties that are tough to be verified}. 
Hence we {\em filter out} those falsified properties.
We obtain around 30 properties for each tested network, as enumerated in \autoref{tab:netdetails}.


\subsection{Network reduction results}
\autoref{tab:reduceSizeRes} shows the size of reduced networks, where the bound propagation methods crown and $\alpha$-crown are used to compute concrete bounds and detect stable neurons.
Here we present the number of neurons in the original network and the average size after reduction (under column ``AvgN") and reduction time (under column ``AvgT") for the two methods.
We have out-of-memory problem when running $\alpha$-crown on network C\_100\_Large, thus we mark the result as ``-".
\begin{table}[!t]
\vspace{-0.5em}
\caption{The average number of ReLU neurons on reduced networks and the mean reduction time in seconds.}
\centering
\def\arraystretch{1.2}
\addtolength{\tabcolsep}{-0.55em}
\begin{tabular}{|l|c||c|c||c|c|}
\hline
\multirow{2}{*}{\textbf{Network}} &\bf Original & \multicolumn{2}{c||}{\bf Reduced (CROWN)}&\multicolumn{2}{c|}{\bf Reduced ($\alpha$-Crown)}\\
\cline{2-6}
&\textbf{\#Neurons}  &
\textbf{AvgN} & \textbf{AvgT(s)}&
\textbf{AvgN} & \textbf{AvgT(s)}\\
\hline
M\_256x6 & 1,536 & 991.77 & 0.14 & 901.10 & 3.10\\
\hline
M\_ConvMed & 5,704 & 2210.77 & 0.21 & 2189.32 & 2.05\\
\hline
 M\_ConvBig & 48,064 & 3250.93 & 0.32 & 3229.76 & 5.03\\
\hline
 M\_SkipNet & 71,650 & 7019.00 & 0.72 & 6796.58 & 7.79\\
\hline
 C\_8\_255Simp &16,634& 2168.13 & 0.33 & 2117.90 & 1.92\\
\hline
C\_WideKW  &  6,244 & 567.06 & 0.28 & 563.47 & 2.08\\
\hline
C\_ConvBig & 62,464 & 6495.00 & 0.39 & 6451.57 & 4.77\\
\hline
C\_Resnet4b &14,436 & 7606.73 & 0.64 & 7449.23 & 10.97\\
\hline
C\_ResnetA & 11,364 & 4654.84 & 0.64 & 4583.06 & 8.08\\
\hline
C\_ResnetB & 11,364 & 4425.90 & 0.60 & 4368.03 & 10.67\\
\hline
C\_100\_Med& 55,460 & 2394.63 & 1.25  &  2352.33  & 12.09 \\
\hline
C\_100\_Large & 286,820  & 7207.29 & 3.50 & -  & - \\
\hline
\end{tabular}
\label{tab:reduceSizeRes}
\end{table}

The table shows that a significant number of neurons could be reduced within a reasonable time budget by leveraging the concrete bounds returned by CROWN.
Therefore, we use CROWN as our bound propagator for the rest of the experiments.
On average, the reduced networks are $10.6 \times$ smaller than the original networks.
%

\begin{figure}[!t]
\centering
\subfigure[Reduction ratio]{
\label{fig:ratio}
\includegraphics[width=0.45\columnwidth]{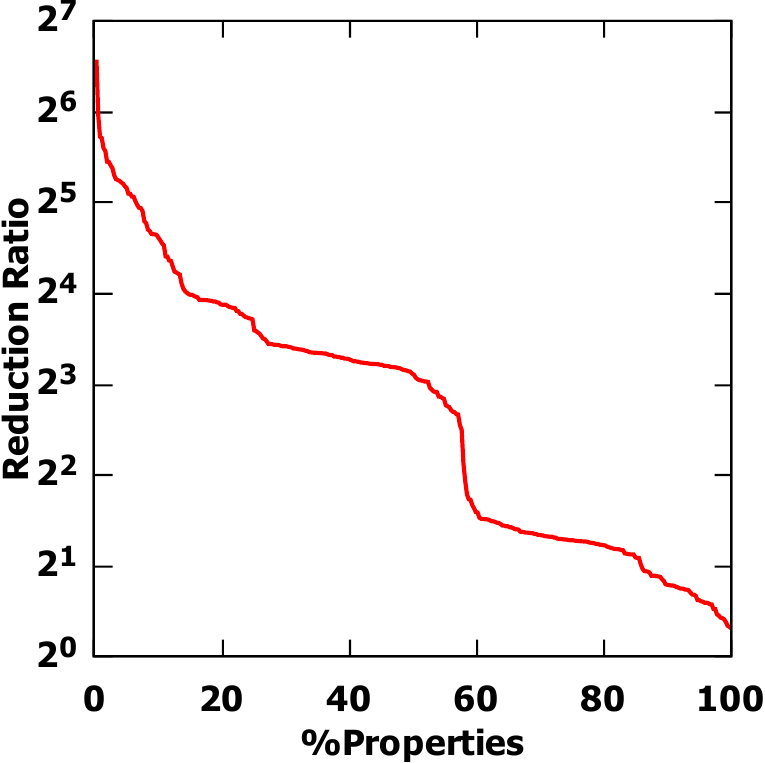}
}
\subfigure[Size distribution]{
\includegraphics[width=0.46\columnwidth]{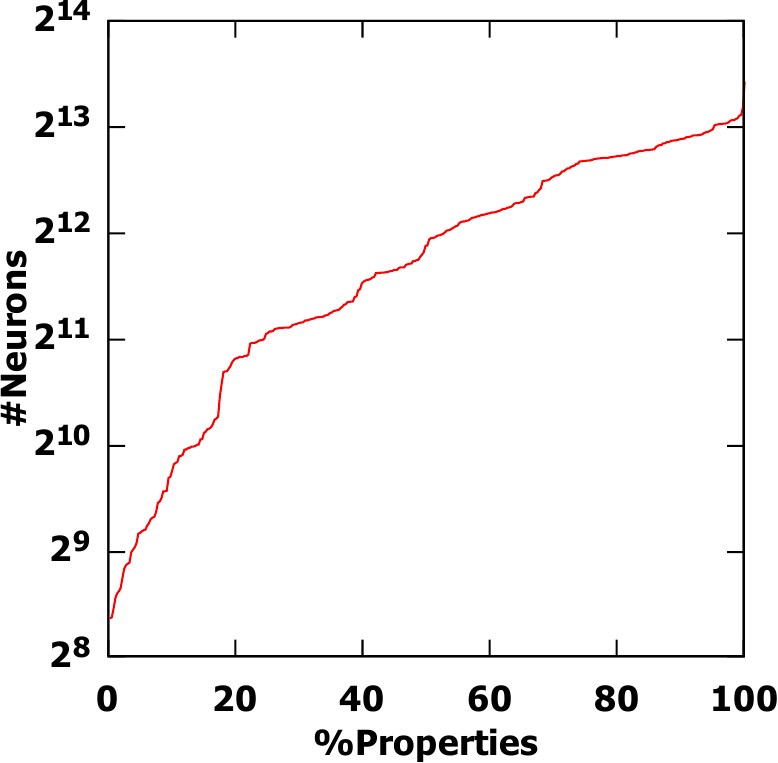}
\label{fig:sizedistribution}
}
\vspace{-0.5em}
\caption{Visualized results of the reduction with CROWN}
\label{fig:reduction}
\end{figure}

%
\autoref{fig:ratio} shows the reduction ratio distribution where each dot $(\alpha, \beta)$ in the figure means that the reduction ratio is greater than $\beta$ on $\alpha$ percent properties.
The reduction ratio can be up to 95 times at the best case and greater than 20 times on 10\% properties.
\autoref{fig:sizedistribution} gives the size distribution of reduced networks.
Each dot $(\alpha, \beta)$ in the figure means the reduced networks have at most $\beta$ ReLU neurons on $\alpha$ percent properties.
We can see that on more than 94\% properties, there are at most 8000 ReLU neurons in the reduced networks.

\subsection{Instantitation on \crown{}}
\label{sec:abcrownRes}
\crown{} is GPU based and the winning verifier in VNNCOMP 2021 \cite{VNNCOMP21} and VNNCOMP 2022 \cite{VNNCOMP22}, the methodology of which is based on linear bound propagation framework and branch-and-bound.
We first instantiate our technique on \crown{}, and we name the new system as \crown{}-R. 
We set the timeout of verification as 300 seconds, and if the tool fails to terminate within the timeout, we deem the result to be inconclusive.
The results are listed in \autoref{tab:abcrownTime}, where we explicitly enumerate the number of timeout properties, the number of verified properties, and the average execution time of \crown{} (column \textbf{\crown{}-O}) and our instantiated system (column \textbf{\crown{}-R}) on the properties where both methods can terminate within timeout.\footnote{When the original method is timeout or fails to execute for all properties, e.g. C\_100\_Med in \autoref{tab:PRIMATime} and M\_ConvBig in \autoref{tab:VerinetTime}, the average time is computed on the properties where our method can terminate within timeout.}

\begin{table}[!hb]
\caption{The results of \crown{} on the original network and the reduced network. The time is the average execution time of the properties where both methods terminate before timeout.}
\vspace{-0.5em}
  \centering
  \def\arraystretch{1.2}
  \addtolength{\tabcolsep}{-0.5em}
  \begin{tabular}{|l|c|c|c|c|c|c|}
    \hline
    \multirow{2}{*}{\textbf{Neural Net}} &
    \multicolumn{3}{c|}{
      \makecell{\textbf{\crown{}-O} \\[-0.2em] {{\footnotesize (on original network)}}}} &
    \multicolumn{3}{c|}{
      \makecell{\textbf{\crown{}-R} \\[-0.2em] {{\footnotesize (on reduced network)}} }}\\

    \cline{2-7}
    &
    \#Timeout & \#Verfied & Time(s) &
    \#Timeout & \#Verfied & Time(s) \\
    \hline
    M\_256x6  & 1 & 29 & 100.53 & 1 & 29 & \textbf{89.81}\\
    \hline
    M\_ConvMed  & 4 &  27 & 48.29 & 2 & \textbf{29} & \textbf{40.41} \\
    \hline
    M\_ConvBig  & 3 & 26 & 43.34 & 1 & \textbf{28} & \textbf{26.16} \\
    \hline
    M\_SkipNet  & 5 & 26 & 38.66 & 2 & \textbf{29} & \textbf{22.35}\\
    \hline
    C\_WideKW  & 1 & 31 & 13.46 & 1 & 31 & \textbf{11.76} \\
    \hline
    C\_8\_255Simp  & 0 & 30 & 19.23  &  0 & 30 & \textbf{16.92} \\
    \hline
    C\_ConvBig  & 1 & 36 & 26.93 & 0 & \textbf{37} & \textbf{19.97} \\
    \hline
    C\_Resnet4b  & 1 & 29 & 39.25 & 1 & 29 & \textbf{30.84} \\
    \hline
    C\_ResnetA  & 0 & 32 & 40.16 & 0 & 32 & \textbf{29.96} \\
    \hline
    C\_ResnetB & 1 & 28 & 28.08 & 0 & \textbf{29} & \textbf{20.54}\\
    \hline
    C\_100\_Med & 4 & 20 & 22.96 & 3 & \textbf{21} & \textbf{9.32}\\
    \hline
    C\_100\_Large & 3 & 21 & 14.29 & 2 & \textbf{22} & \textbf{5.68}\\
    \hline
  \end{tabular}
  \label{tab:abcrownTime}
\end{table}

From the result, we observe that \crown{}-R could verify more tough properties that have failed to be verified within 300 seconds in \crown{}-O.
This indicates that our reduction pre-processing does not only benefit those easy verification problems but also helps verify more difficult properties within a decent time.
In general, \crown{}-R verifies 11 more properties and boosts the efficiency of \crown{}-O with average $1.52 \times$ speedup on all 12 networks. 
The average is computed across the networks where the speedup for each network is calculated by $\frac{\text{the reported time on the original network}}{\text{the reported time on the reduced network}}$.

In addition, the performance of REDNet is affected by the network reduction ratio.
For example, \crown{}-R only has average 1.12 speedup on M\_256$\times$6 whose reduction ratio is only 1.55, while \crown{}-R can have average 2.52$\times$ speedup on C\_100\_large whose mean reduction ratio is 39.80.



\subsection{Instantitation on PRIMA}
\label{sec:PRIMAexperiments}
PRIMA \cite{DBLP:journals/pacmpl/MullerMSPV22} is one of the state-of-the-art incomplete verification tools. It introduces a new convex relaxation method that considers multiple ReLUs jointly in order to capture the correlation between ReLU neurons in the same layer.
Furthermore, PRIMA leverages LP-solving or MILP-solving to refine individual neuron bounds within a user-configured timeout.
Note that PRIMA stores the connection between neurons in two ways: 1. Dense expression, which encodes the fully-connected computation in a fully-connected layer; 2. Sparse expression, that only keeps the non-zero coefficients and the indexes of preceding neurons of which the corresponding coefficients are non-zero (e.g. convolutional layer).
As some affine connections between layers in our reduced network contain many zero elements (since we introduce the identity matrix in the newly constructed connection), we elect to record them as sparse expressions in the instantiated PRIMA (abbreviated as PRIMA-R). 
\begin{table}[!ht]
\caption{The experiment results of PRIMA on the original network and the reduced network. When PRIMA-O fails to
execute or times out for all the properties, e.g. M\_SkipNet or C\_100\_Med, the average time is computed on the properties where our method can terminate within the timeout.}
  \centering
  \def\arraystretch{1.2}
  \addtolength{\tabcolsep}{-0.6em}
  \vspace{-0.5em}
  \begin{tabular}{|l|c|c|c|c|c|c|}
    \hline
    \multirow{2}{*}{\textbf{Neural Net}} &
    \multicolumn{3}{c|}{
      \makecell{\textbf{PRIMA-O} \\[-0.2em] {{\footnotesize (on original network)}}}} &
    \multicolumn{3}{c|}{
      \makecell{\textbf{PRIMA-R} \\[-0.2em] {{\footnotesize (on reduced network)}} }}\\

    \cline{2-7}
    &
    \#Unknown & \#Verfied & Time(s) &
    \#Unknown & \#Verfied & Time(s) \\
    \hline
    M\_256x6  & 30 & 0 & 299.94 & 30 & 0 & \textbf{281.77}\\
    \hline
    M\_ConvMed  & 23 & 8 & 244.45 & 22 & \textbf{9} & \textbf{196.83}\\
    \hline
    M\_ConvBig  & 23 & 6 & 352.71 & 23 & 6 & \textbf{75.13}\\
    \hline
    M\_SkipNet & - & - & - & 30 & \textbf{1} & \textbf{432.08}\\
    \hline
    C\_WideKW  & 3 & 29 & 53.80 & 3 & 29 & \textbf{10.21} \\
    \hline
    C\_8\_255Simp  & 30 & 0 & 329.94 & 27 & \textbf{3} & \textbf{255.97}\\
    \hline
    C\_ConvBig & 32 & 5 & 227.81 & 23 & \textbf{14} & 282.40 \\
    \hline
    C\_Resnet4b  & 25 & 5 & 912.37 & 25 & 5 & \textbf{757.86}\\
    \hline
    C\_ResnetA  & 28 & 4 & 537.36 & 28 & 4 & \textbf{459.67} \\
    \hline
    C\_ResnetB & 25 & 4 & 486.68 & 23 & \textbf{6} & \textbf{416.12}\\
    \hline
    C\_100\_Med & 24 & 0 & TO & 14 & \textbf{10} & \textbf{135.97}\\
    \hline
    C\_100\_Large & 24 & 0 & TO & 13 & \textbf{11} & \textbf{243.92}\\
    \hline
  \end{tabular}
  \label{tab:PRIMATime}
\end{table}

The comparison results are given in \autoref{tab:PRIMATime}, and we set a 2000 seconds timeout for each verification query as PRIMA runs on the CPU and usually takes a long execution time for deep networks. 
Note that PRIMA returns segmentation fault for M\_SkipNet, thus the results are marked as ``-"; PRIMA times out for all properties of C\_100\_Med and C\_100\_Large, hence marked as ``TO''.
Note that there are some cases where PRIMA-R runs slower than PRIMA-O, e.g., for network C\_ConvBig. 
This happens because PRIMA conducts refined verification by pruning the potential adversarial label one by one within a certain timeout. 
Once an adversarial label fails to be pruned within the timeout, PRIMA returns unknown immediately without checking the rest of the adversarial labels.
In PRIMA-R, we could prune those failed labels that previously timed out in PRIMA-O, thus continuing the verification process, which may take more overall time.
But accordingly, we gain significant precision improvement, e.g. PRIMA-R can verify 9 more properties on C\_ConvBig.

On average, PRIMA-R gains $1.99 \times$ speedup than PRIMA-O and verifies 60.6\% more images, which indicates the strength of REDNet to improve both efficiency and precision.

\subsection{Instantiation on VeriNet}
\label{sec:verinetres}
VeriNet \cite{VerinetSystem} is the state-of-the-art complete symbolic interval propagation based toolkit.
It is the second-place winner in VNNCOMP 2021.
Similarly, we present the result of the original VeriNet tool under the column VeriNet-O at \autoref{tab:VerinetTime}; the instantiation of REDnet on VeriNet is named VeriNet-R.
The time reported is the average execution time on properties where both VeriNet-O and VeriNet-O terminate within 300 seconds of timeout.
We use a free FICO Community license for the XPress solver called by VeriNet.
Thus, we only consider 8 networks which fit the limits of the license.

\begin{table}[!ht]
\caption{The experiment results of VeriNet. The time is the average execution time of the properties where both methods terminate before timeout. When VeriNet-O fails to execute,
e.g. M\_ConvBig, the average time is
computed on the properties where our method can terminate within the timeout.}
  \centering
  \def\arraystretch{1.2}
  \addtolength{\tabcolsep}{-0.6em}
 \vspace{-0.5em}
  \begin{tabular}{|l|c|c|c|c|c|c|}
    \hline
    \multirow{2}{*}{\textbf{Neural Net}} &
    \multicolumn{3}{c|}{
      \makecell{\textbf{VeriNet-O} \\[-0.2em] {{\footnotesize (on original network)}}}} &
    \multicolumn{3}{c|}{
      \makecell{\textbf{VeriNet-R} \\[-0.2em] {{\footnotesize (on reduced network)}} }}\\

    \cline{2-7}
    &
    \#Timeout & \#Verfied & Time(s) &
    \#Timeout & \#Verfied & Time(s) \\
    \hline
    M\_256x6  & 27 & 3 & 52.94 & 27 & 3 & \textbf{47.78}\\
    \hline
    M\_ConvMed  & - & - & - & 19 & \textbf{12} & \textbf{23.96}\\
    \hline
    M\_ConvBig  & - & - & - & 23 & \textbf{6} & \textbf{48.81}\\
    \hline
    C\_WideKW  & 3 & 29 & 27.12 & 2 & \textbf{30} & \textbf{22.74}\\
    \hline
    C\_8\_255Simp  & 10 & 20 & 63.26 & 10 & 20 & \textbf{52.56}\\
    \hline
    C\_ResnetA  & 25 & 7 & 129.49 & 25 & 7 & \textbf{83.35} \\
    \hline
    C\_ResnetB & 21 & 8 & 116.80 & 20 & \textbf{9} & \textbf{74.98}\\
    \hline
    C\_100\_Med & 10 & 14 & 69.71 & 9 & \textbf{15} & \textbf{21.15}\\
    \hline
  \end{tabular}
  \label{tab:VerinetTime}
\end{table}

In general, VeriNet-R can verify 25.9\% more properties than VeriNet-O.
On average, VeriNet-R can be $1.65\times$ faster than VeriNet-O. 
Additionally, the result in \autoref{tab:VerinetTime} marked with ``-'' means that the neural networks M\_ConvMed and M\_ConvBig are not supported by VeriNet.
This shows that network reduction can improve the availability of VeriNet.

\subsection{Overall comparison in visualized figures}
\label{sec:overallCompare}
%
\begin{figure}
\centering
\subfigure[\crown{}-R/-O]{
\label{fig:overall-crown}
\includegraphics[width=0.45\columnwidth]{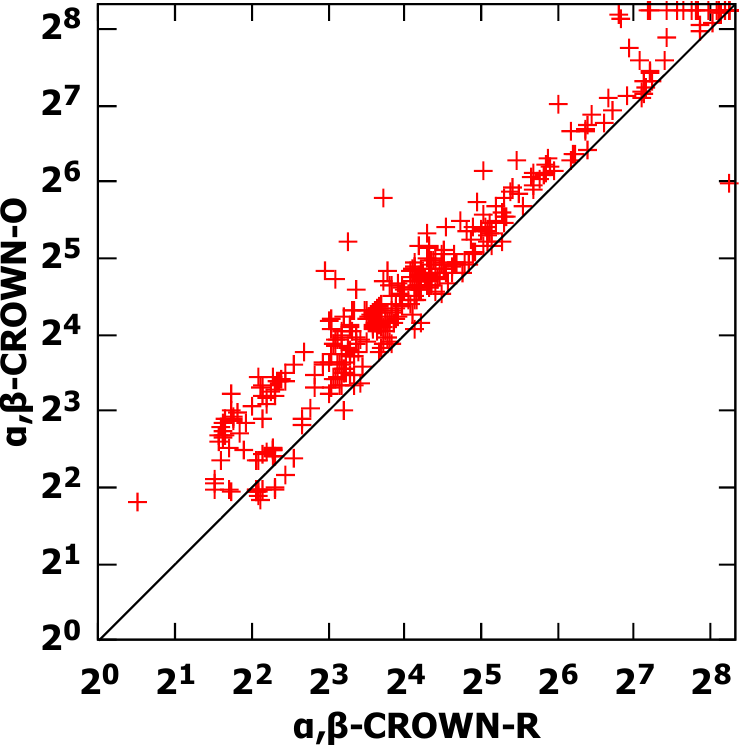}
}
\subfigure[\crown{} time distribution]{
\label{fig:overall-crown-time}
\includegraphics[width=0.46\columnwidth]{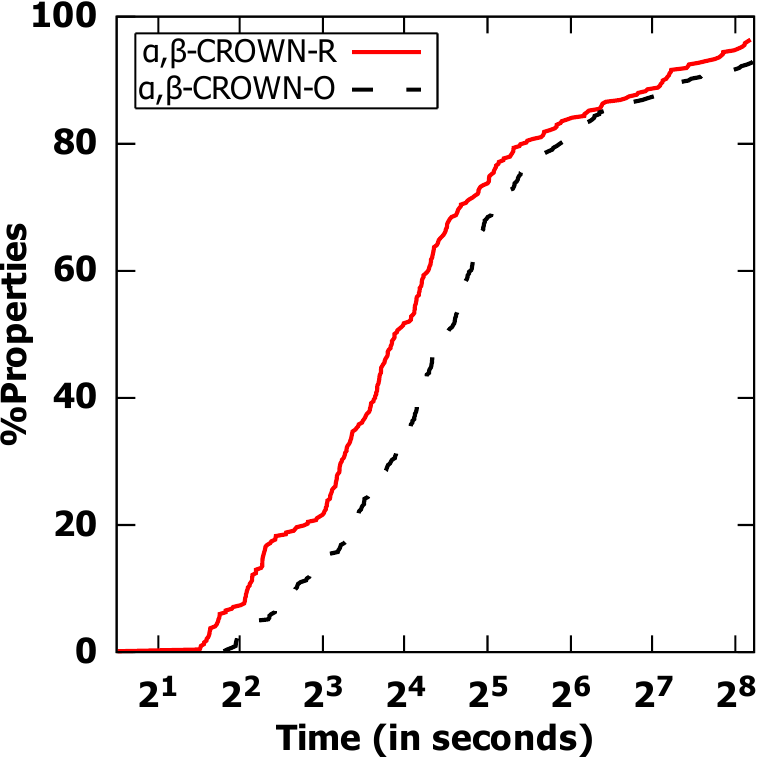}
}
\vspace{-0.5em}
\subfigure[PRIMA-R v.s. PRIMA-O]{
\label{fig:overall-prima}
\includegraphics[width=0.47\columnwidth]{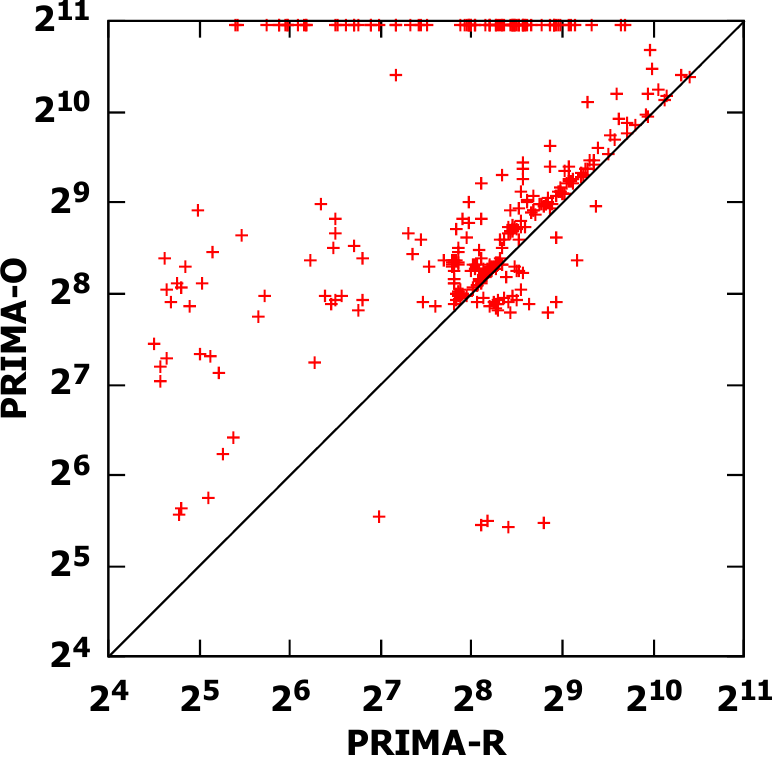}
}
\subfigure[VeriNet-R v.s. VeriNet-O]{
\label{fig:overall-verinet}
\includegraphics[width=0.45\columnwidth]{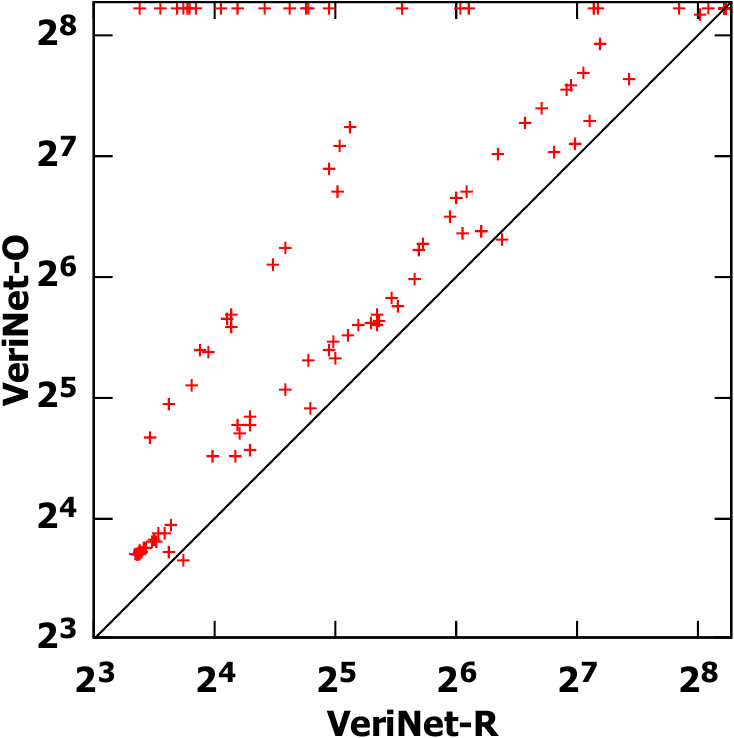}
}
\vspace{-0.5em}
\subfigure[\crown{} and VeriNet results]{
\label{fig:abcveriEffect}
\includegraphics[width=0.46\columnwidth]{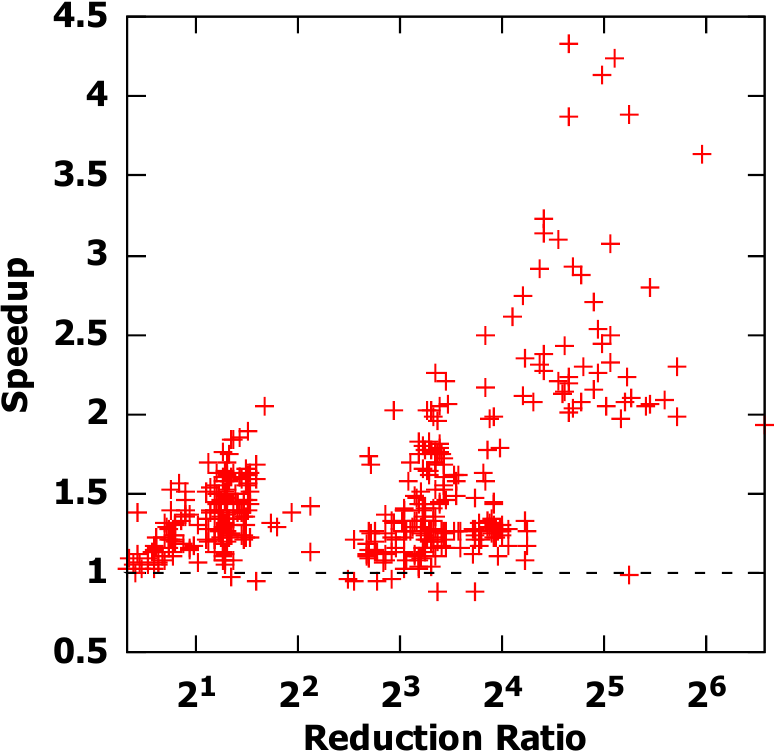}
}
\subfigure[PRIMA speedup results]{
\includegraphics[width=0.456\columnwidth]{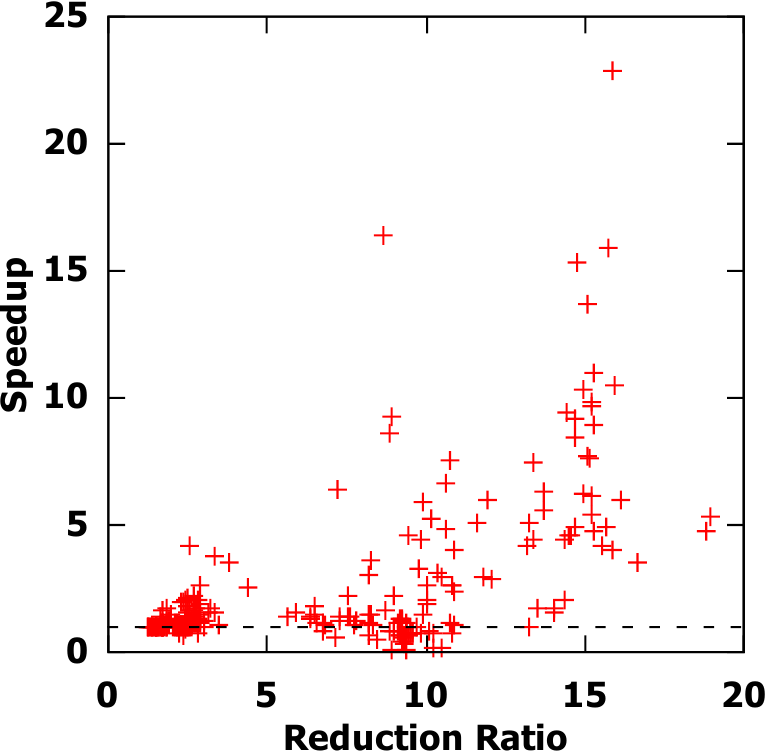}
\label{fig:prima-size-speed}
}
\vspace{0.5em}
\caption{Visualized comparison results.}
\label{fig:overall}
\end{figure}

\autoref{fig:overall} gives a visualized comparison between the verification tools on the original network and those on REDNet.
\autoref{fig:overall-crown}, \autoref{fig:overall-prima} and \autoref{fig:overall-verinet} shows the execution time of the verification tools on all tested properties.
Each dot in the figures denotes a property, and both the x-axis and y-axis indicate execution time in seconds.
The result of a verification tool on an unsupported network is regarded as timeout.
Then \autoref{fig:overall-crown-time} shows the execution time distribution of \crown{}-R and \crown{}-O, where each position $(\alpha, \beta)$ denotes that the tool can verify $\beta$ percent properties in $\alpha$ seconds.
For example, by setting the time limit to 10 seconds, \crown{}-R can verify 32.9\% properties, and \crown{}-O only verifies 18.3\% properties.

On most properties, the verification tools on REDNet are faster than the tools on the original network.
Despite its generality, REDNet may achieve marginal effectiveness on certain tools or benchmarks due to the following factors:
\begin{itemize}
\item The reduction ratio affects the subsequent verification acceleration. A less significant reduction ratio plus the reduction cost could cause marginal overall speedup.
\autoref{fig:abcveriEffect} and \autoref{fig:prima-size-speed} depict the effect of the reduction ratio on the speedup gained.
Despite other factors affecting the final speedup, there is a general trend that a significant reduction ratio leads to better speedup, which may cause superb effectiveness on networks C\_100\_Med and C\_100\_Large.
\item Different tools may use distinct bound propagation methods, which have different degrees of dependency on the network size. 
PRIMA deploys DeepPoly whose time complexity depends on $N^3$ where each layer has at most $N$ neurons \cite{DBLP:conf/mlsys/MullerS0PV21}; as such reduction in network size can lead to better performance.
\crown{}, on the other hand, uses $\beta$-crown. 
$\beta$-crown is used to generate constraints of output neurons defined over preceding layers until the input layer.
Thus, the number of constraints does not vary, and the number of intermediate neurons can only affect the number of variables that appear in the constraint; as such, deployment of REDNET may reap a marginal effect in speedup on \crown{} compared to PRIMA.
For VeriNet, it uses symbolic interval propagation to generate constraints of intermediate and output neurons defined over the input neurons.
Thereby intermediate neuron size only affects the number of constraints while the number of defined variables in the constraint is fixed as the input dimension. 
Hence, REDNet could be less effective on VeriNet compared to PRIMA in general.
 \item Some layer types (e.g. Conv) may compute faster than fully-connected layers; since our method transforms these layers into fully-connected layers before performing network reduction, its efficiency may not be that significant as compared to the original layers. On the other hand, it is worth-noticing that the use of fully-connected layers improves the availability of existing tools.  
\item \crown{} and VeriNet are branch-and-bound based and they generate sub-problems from their respective branching heuristics, which are dependent on the original network structures. The REDNet changes the network structure, and hence the heuristic can generate different sub-problems. 
This may affect the performance.
\end{itemize}

We conclude empirically that REDNet has better performance (significant speedup or much more properties verified) on large networks, i.e. networks with more than 40k ReLU neurons. 
On the large networks, the average speedup of \crown{}-R is 1.94$\times$, and the average speedup of VeriNet-R is 3.29$\times$;
%
and PRIMA-R verifies 42 properties while PRIMA-O only verifies 11 properties.

\subsection{Support of other verifiers for the benchmarks}
As can be seen from \autoref{sec:PRIMAexperiments}, PRIMA fails to analyze M\_SkipNet because it does not support its network architecture. 
However, with the introduction of REDNet, which is constructed as a fully-connected neural network, PRIMA is now able to verify M\_SkipNet. 
%
A similar improvement happens to VeriNet.
Therefore, our REDNet not only speeds up the verification process but also allows existing tools to handle network architectures that are not supported originally.

To further testify that the reduced network adds support to existing verification tools, we select four tools - Debona\cite{DebonaSystem}, Venus\cite{VenusSystem}, Nnenum\cite{NnenumSystem}, PeregriNN\cite{PeregriNNSystem} - from VNNCOMP2021/2022 that only support limited network architectures.
We select one representative verification property for each of our tested networks to check if the four designated tools can support the networks.

\begin{table}[!ht]
\caption{The networks supported by existing verification tools. A fully black circle indicates both the original and the reduced networks are supported. A right-half black circle indicates that the tool supports only the reduced network.}
\centering
\def\arraystretch{1.2}
\addtolength{\tabcolsep}{-0.6em}
\begin{tabular}{|l|c|c|c|c|c|}
\hline
\multirow{2}{*}{\bf Networks(12)}& \multicolumn{4}{c|}{\bf Tools} \\
\cline{2-5}
&\multicolumn{1}{c|}{ Venus} & \multicolumn{1}{c|}{Debona} &\multicolumn{1}{c|}{Nnenum}& \multicolumn{1}{c|}{PeregriNN}\\
\hline
M\_256x6 & \CIRCLE & \CIRCLE& \CIRCLE& \CIRCLE \\ 
M\_ConvMed & \RIGHTcircle & \RIGHTcircle & \RIGHTcircle & \RIGHTcircle\\ 
M\_ConvBig & \RIGHTcircle & \RIGHTcircle & \RIGHTcircle & \RIGHTcircle\\ 
M\_SkipNet & ReLU-error & \RIGHTcircle & \RIGHTcircle & \RIGHTcircle\\ 
C\_WideKW & \CIRCLE & \RIGHTcircle & \CIRCLE& \CIRCLE\\ 
C\_8\_255Simp & \CIRCLE & \RIGHTcircle & \CIRCLE & \RIGHTcircle\\ 
C\_ConvBig & \RIGHTcircle& \RIGHTcircle & \RIGHTcircle& \RIGHTcircle\\ 
C\_Resnet4b & \RIGHTcircle &\RIGHTcircle & \RIGHTcircle & \RIGHTcircle\\ 
C\_ResnetA & \RIGHTcircle& \RIGHTcircle & \RIGHTcircle& \RIGHTcircle\\ 
C\_ResnetB & \RIGHTcircle& \RIGHTcircle & \RIGHTcircle& \RIGHTcircle \\ 
C\_100\_Med & \RIGHTcircle & \RIGHTcircle & \RIGHTcircle & \RIGHTcircle\\ 
C\_100\_Large & \RIGHTcircle & \RIGHTcircle & \RIGHTcircle & \RIGHTcircle\\ 
\hline
\end{tabular}
\label{tab:benchSupport}
\end{table}

We present the results in \autoref{tab:benchSupport} where we color the left half of the circle black to indicate that the original network is supported by the tool (and white otherwise); we also color the right half of the circle black if the reduced network is supported by the tool (and white otherwise.)
In general, the black color implies the network is \emph{supported} and the white color implies the network is \emph{not supported}.
Note that Venus does not support networks whose output layer is a ReLU layer; therefore, it cannot be executed for both the original and the reduced network for M\_SkipNet.
%
These results boost our confidence that our constructed REDNet not only accelerates the verification but also {\em produces a simple neural network architecture that significantly expands the scope of neural networks which various tools can handle.}

\section{Discussion}
\label{sec:discussion}
We now discuss the limitation of our work.

{\em Supported layer types.} As described in \autoref{sec:simplyMethod}, our reduced neural network contains only Affine layers (e.g. GEMM layers) and ReLU layers, therefore we could only represent non-activation layers that conduct linear computation.
For example, an \emph{Add} layer that takes layer $\alpha$ and layer $\beta$ conducts linear computation as the output is computed as $\alpha+\beta$.
A \emph{Convolutional} layer conducts linear computation as well as it only takes one input layer and the other operands are constant weights and bias.
However, we couldn't support a \emph{Multiplication} layer if it takes layer $\alpha$ and layer $\beta$ and computes $\alpha \times \beta$ as the output.
For future work, we will explore the possibility of handling more non-linear computations.

{\em Floating-point error.}
%
As presented in \autoref{theorem:equivalence}, our reduction process preserves the input-output equivalence of the original network in the real-number domain.
However, like many existing verification algorithms \cite{DBLP:conf/nips/SinghGPV19, DBLP:journals/pacmpl/MullerMSPV22, zhang2022general, AlphaBetaCrownSystem} that use floating-point numbers when conducted on physical machines, our implementation involves floating-point number computation, thus inevitably introducing floating-point error.
The error could be mitigated by deploying float data type with higher precision during implementation.

\section{Related Work}
\label{sec:relatedwork}
Theoretically, verifying deep neural networks with ReLU functions is an NP-hard problem \cite{DBLP:journals/corr/KatzBDJK17}. 
Particularly, the complexity of the problems grows with a larger number of {\em nodes} in the network. 
Therefore, with the concern of scalability, many works have been proposed by over-approximating the behavior of the network. 
This over-approximation can be conducted by abstract interpretation techniques \cite{DBLP:conf/cav/PulinaT10, DBLP:conf/sp/GehrMDTCV18, DBLP:journals/pacmpl/SinghGPV19}; or to soundly approximate the network with {\em fewer nodes} \cite{DBLP:conf/sas/SotoudehT20, DBLP:conf/nips/PrabhakarA19, DBLP:journals/corr/abs-2305-01932, DBLP:journals/corr/abs-1910-14574}.

In detail, abstract interpretation-based methods over-approximate the functionality of each neuron with an abstract domain, such as box/interval \cite{DBLP:conf/cav/PulinaT10}, zonotope \cite{DBLP:conf/sp/GehrMDTCV18} or polyhedra \cite{DBLP:journals/pacmpl/SinghGPV19}.
These methods reason over the original neural networks without changing the number of neurons in the test network.

On the contrary, reduction methods in \cite{DBLP:conf/sas/SotoudehT20, DBLP:conf/nips/PrabhakarA19, DBLP:journals/corr/abs-2305-01932, DBLP:journals/corr/abs-1910-14574} reduce the number of neurons in a way that over-approximates the original network's behavior.
%
%
However, such over-approximation would jeopardize completeness when instantiated on complete methods.
On the contrary, our reduction method captures the {\em exact} behavior of the network without approximation error.
Therefore REDNet could be instantiated on complete tools and even verify more properties given the same timeout.
Furthermore, REDNet could handle various large networks where the previous work \cite{DBLP:journals/corr/abs-2305-01932} only evaluated one large-scale network (the C\_ConvBig in our benchmark) that was reduced to 25\% of the original size with a very small perturbation $\epsilon=0.001$; whereas we could reduce it to just 10\% with $\epsilon\approx 0.0078$ (properties from VNN competition 2022).
We remark that the smaller perturbation, the more reduction we could gain.
Other related tools in \cite{DBLP:conf/nips/PrabhakarA19, DBLP:journals/corr/abs-1910-14574} were only evaluated with ACAS Xu networks with a very small input dimension and network sizes, making it challenging for us to make any meaningful comparison. 
Last but not least, the reduced networks designed in \cite{DBLP:conf/sas/SotoudehT20, DBLP:conf/nips/PrabhakarA19} use intervals or values in an abstract domain to represent connection weights.
Such specialized connections require implementation support if instantiated on existing verification methods.
But we export REDNet as a fully-connected network in ONNX, which is an open format that is widely accepted.
This makes our REDNet a versatile tool that could also be combined with existing reduction methods to lessen the network size even further as they all apply to fully-connected networks.

REDNet could benefit various verification techniques, such as branch-and-bound based methods \cite{DBLP:journals/jmlr/BunelLTTKK20, DBLP:conf/iclr/XuZ0WJLH21, DBLP:conf/nips/BunelTTKM18, DBLP:conf/iclr/FerrariMJV22}.
The key insight of the branch-and-bound method is to divide the original verification problem $P$ into subdomains/subproblems by splitting neuron intervals.
For example, one can bisect the input neuron interval such that the input space is curtailed in each subdomain; or to split an unstable ReLU neuron $y$ (whose input value can be both negative and positive) at the point 0, thereby $y$ will be stably activated or stably deactivated in the subdomains.
Our {\em network reduction} technique, once applied at the beginning of branch-and-bound based methods, will help generates easier subproblems based on a small-sized network, thus accelerating the whole analysis process without sacrificing verification precision.

Furthermore, the reduced network could accelerate abstract refinement based processes like PRIMA \cite{DBLP:journals/pacmpl/MullerMSPV22}, where it encodes the network constraints and resolves individual neuron bounds. 
As REDNet contains fewer neurons and connections, the solving process involves a smaller set of constraints, which leads to overall speedup. 

\section{Conclusion}
In this work, we propose the {\em neural network reduction} technique, which constructs a reduced neural network with fewer neurons and connections while capturing the \emph{same} behavior as the original tested neural network.
In particular, we provide formal definitions of stable ReLU neurons and deploy the state-of-the-art bound propagation method to detect such stable neurons and \emph{remove} them from the neural network in a way that preserves the network functionality.
We conduct extensive experiments over various benchmarks and state-of-the-art verification tools. 
The results on a large set of neural networks indicate that our method can be instantiated on different verification methods, including \crown{}, VeriNet and PRIMA, to expedite the analysis process further. 

We believe that our method is an efficient pre-processing technique that returns a functionally-equivalent reduced network on which the same verification algorithm runs faster, correspondingly enhancing the efficiency of existing verification methods for them to answer tough verification queries within a decent time budget. 
Moreover, the simplified network architectures in REDNets empower existing tools to handle a wider range of networks they could not support previously.

\section*{Acknowledgment}
This research is supported by a Singapore Ministry of Education Academic Research Fund Tier 1 T1-251RES2103. 
The second author is supported by NUS grant T1 251RES2219. 


\bibliographystyle{IEEEtran}
\bibliography{refe}

\end{document}